\documentstyle[preprint,aps,floats,epsfig]{revtex}
\newcommand{\ca}{c^{\phantom{\dagger}}}
\newcommand{\cc}{c^\dagger}
\newcommand{\da}{d^{\phantom{\dagger}}}
\newcommand{\dc}{d^\dagger}
\newcommand{\be}{\begin{equation}}
\newcommand{\ee}{\end{equation}}
\newcommand{\bea}{\begin{eqnarray}}
\newcommand{\eea}{\end{eqnarray}}
\newcommand{\ben}{\begin{eqnarray*}}
\newcommand{\een}{\end{eqnarray*}}
\begin{document}
\title{Anderson localization in bipartite lattices}
\author{Michele Fabrizio$^{1,2}$ and Claudio Castellani$^3$}
\address{$^1$ Istituto Nazionale di Fisica della Materia,}
\address{and International School for Advanced Studies (SISSA),
via Beirut 2-4, I-34014 Trieste, Italy,}
\address{$^2$ International Center for Theoretical Physics 
(ICTP), Trieste, Italy}
\address{$^3$ Istituto Nazionale di Fisica della Materia and 
Universit\`a degli Studi di Roma ``La Sapienza'', 
Piazzale Aldo Moro 2, I-00185, Roma, Italy} 
\maketitle
\begin{abstract}
We study the localization properties of a disordered tight-binding  
Hamiltonian on a generic bipartite lattice close to the band center. 
By means of a fermionic replica trick method, we derive the effective 
non-linear $\sigma$-model describing the diffusive modes, 
which we analyse by using the 
Wilson--Polyakov renormalization group. In addition to the 
standard parameters which define the non-linear $\sigma$-model, namely 
the conductance and the external frequency, a new parameter enters, 
which may be related to the 
fluctuations of the staggered density of states.   
We find that, when both the regular hopping and the 
disorder only couple one sublattice to the other, 
the quantum corrections to 
the Kubo conductivity vanish at the band center, thus implying the 
existence of delocalized states. In two dimensions, the RG equations 
predict that the conductance flows to a finite value, while
both the density of states and the staggered density of states fluctuations 
diverge. In three dimensions, we 
find that, sufficiently close to the 
band center, all states are extended, independently of the 
disorder strength. 
We also discuss the role of various symmetry breaking terms, 
as a regular hopping between same sublattices, or an on-site disorder.  
\end{abstract}
\section{Introduction}

An interesting and still debated issue in the physics of the Anderson's 
localization concerns the existence of delocalized states in 
dimensions $d\leq 2$, the conditions under which they appear, and 
their properties. This problem, which is, for instance, of relevance 
in the theory of the integer quantum Hall effect\cite{QHE},   
got recently a   
renewed interest after evidences 
of a metal--insulator transition in two dimensions have been 
discovered\cite{MIT}. 
 
One of the cases in which localization 
does not occur in any dimension is at the band center 
energy of a tight binding model 
on a bipartite lattice, when both the regular hopping and the disorder 
only couple one sublattice to the other, i.e. in the so 
called {\sl two sublattice model}\cite{T&C,E&R,Wegner}.  
Although this is not a common physical situation, 
its consequences are surprising, and seem to escape 
any quasi-classical interpretation, which, on the contrary,  
provides simple physical explanations of other delocalization mechanisms
\cite{Bergmann}. 
Already in 1976, Theodorou and Cohen\cite{T&C} realized that a 
one dimensional tight binding model with nearest neighbor random hopping 
has a single delocalized state at the band center 
(see also Ref.\onlinecite{E&R}). 
Afterwards, Wegner\cite{Wegner} and Oppermann and Wegner\cite{O&W} 
showed that a delocalized state indeed exists under the 
above conditions in any dimension, 
within a large--$n$ expansion, being $n$ the 
number of orbitals per site.  
Later on, Wegner and Gade\cite{W&G} argued that these models 
correspond to a particular class of non-linear $\sigma$-models for 
matrices in the zero replica limit. They were able to show 
that the quantum corrections to the $\beta$-function 
which controls the scaling behavior 
of the conductance vanish at the band center at all orders in the 
disorder strength.
thus implying a metallic behavior at this value of the chemical potential. 
Moreover, they showed that, contrary to the standard case, the  
$\beta$-function of the density of states is finite.
These results were based upon the non-linear $\sigma$-model 
derived by Gade\cite{Gade} by means of a boson-replica 
trick method, in a particular two sublattice Hamiltonian 
with broken time reversal invariance.

More recently, the model without time reversal invariance 
has got a renewed interest for its 
implications in different physical contexts, for instance models 
with non-Hermitean stochastic operators, or {\sl random flux} models 
in two dimensions (see e.g. Refs.\onlinecite{Fukui,Altland,Ludwig}).  

In this paper, we present an analysis of a generic disordered tight-binding 
Hamiltonian on a 
generic bipartite lattice. The starting model is therefore of 
quite general validity, also describing systems with time reversal 
invariance, and reduces in particular cases to the 
model discussed by Gade\cite{Gade}, or, in the honeycomb lattice, to 
models of Dirac fermions\cite{Ludwig}, or, finally, to 
{\sl random flux} models\cite{Fukui,Altland}. 
By means of a fermionic-replica trick method, we derive the generic  
non-linear $\sigma$-model describing the diffusive modes, 
which we analyse by the Renormalization Group (RG). 
Needless to say, the effective model belongs to the same  
class of non-linear $\sigma$-models identified by Wegner and Gade, 
demonstrating once more the universality of this description in the 
theory of Anderson localization\cite{Wegnersigma}. 

Since the work is quite technical, we prefer to 
give in the following section a short summary of the main results. 

\subsection{Summary of the main results}
\label{Summary}

In this section we shortly present the main results,  
with particular emphasis to the connections and differences with the standard 
theory of the Anderson's localization.  

We consider a generic bipartite lattice and work with a unit cell which 
contains two sites from opposite sublattices. The  
Pauli matrices $\sigma$'s act on the two components of the wavefunction,  
corresponding to the two sites within each unit cell.  
In this lattice, we study   
a disordered tight-binding Hamiltonian which has the peculiar 
property of {\sl involving only both    
Pauli matrices $\sigma_1$ and $\sigma_2$}. 
In other words, $H$ satisfies the conditions 
\be
\left\{H,\sigma_3\right\}=0,\;\;\;
\left\{H,\sigma_1\right\}\not = 0,\;\;\;  
\left\{H,\sigma_2\right\}\not = 0,
\label{Summary:condition}
\ee
where $\{\dots,\dots\}$ indicates the anticommutator.
By means of a path-integral approach within a fermionic replica trick method, 
we find that the low-energy diffusive modes 
at the band center, $E=0$, can be represented by the 
non-linear $\sigma$-model 
\bea
S[U] &=& \frac{2\pi\sigma_{xx}}{16} \int dR\; 
Tr\left[ \vec{\nabla}U(R)^{-2} \cdot \vec{\nabla}U(R)^2\right]
\nonumber\\
&-& \frac{2\pi}{ 32}\frac{\Pi}{2}\int dR\; 
\left\{Tr\left[U(R)^{-2}\, \vec{\nabla}U(R)^2\right]\right\}^2,
\label{Summary:NLSM}
\eea
where $U(R)$ is unitary and belongs to the 
coset space U($4m$)/Sp($2m$), being $m$ the number of replicas. 
At finite energy $E\not =0$, the symmetry of $U(R)$ gets reduced to 
Sp($2m$)/Sp($m$)$\times$ Sp($m$), as in the standard case\cite{EL&K}. 
The enlarged symmetry is accompanied by new diffusive modes which appear 
in the retarded-retarded and advanced-advanced channels, which are 
instead massive in the standard case.  
In (\ref{Summary:NLSM}), $\sigma_{xx}$ is the Kubo conductivity 
(in units of $e^2/\hbar$) in the 
Drude approximation. We find that the new coupling constant, $\Pi$, 
is proportional to the fluctuations of the staggered density of states, 
i.e. to the following correlation function 
\be
\frac{1}{V}
\sum_{RR'} {\rm e}^{-iq(R-R')}
(\overline{\langle\rho_s(E,R)\rangle\langle\rho_s(E,R')\rangle},
\ee
where the bar indicates the impurity average, and $\rho_s(E,R)$ 
is the staggered density of states at energy $E$,
\[
\rho_s(E,R) = \sum_n \phi_n(R)^\dagger \sigma_3 \phi_n(R) 
\delta(E-\epsilon_n),
\]
being $\phi_n(R)$ the two-component eigenfunction of energy $\epsilon_n$.

The structure of the above action was derived by 
Gade and Wegner\cite{W&G} for a particular Hamiltonian. Here, 
we derive it for a generic 
bipartite lattice and random hopping. 
Moreover, we provide a simple physical interpretation of $\Pi$.

Going back to (\ref{Summary:NLSM}), Gade and Wegner\cite{W&G} 
gave a beautiful proof, 
based just on symmetry considerations, that the quantum corrections to 
the $\beta$-function of $\sigma_{xx}$ vanish in the zero replica limit. 
In Appendix \ref{App:Symmetries}, we outline how their proof works in 
our case, which is sligthly, but not qualitatively, different from the 
U(N)/SO(N) case they have considered. Essentially, one can show that 
the action (\ref{Summary:NLSM}) posseses an invariant coupling 
$\sigma_{xx}+m\Pi$, which, in the $m\to 0$ limit, implies that 
$\sigma_{xx}$ is not renormalized, apart from its bare dimensions. 

On the contrary, both the density of states and $\Pi$ have  
non vanishing $\beta$-functions. In $d=1$, the system flows to strong 
coupling, hence we can not access the asymptotic infrared behavior. 
Nevertheless, the starting flow of the running variables indicates that 
the density of states diverges. 
In $d=2,3$, the system flows to weak coupling, hence we can safely 
assume that the infrared behavior is captured by the RG equations. 
Indeed, in two dimensions, the density of 
states diverges at $E=0$, while, in $d=3$, it saturates to a finite 
value, although exponentially increased in 
$1/\sigma_{xx}$. 
Moreover, $\Pi$ has an anomalous behavior in $d=2$, where it is 
predicted to diverge logarithmically. 
We explicitly estimate how these quantities 
behave as $E\to 0$,  
by means of a two-cutoff scaling approach, 
as discussed by Gade\cite{Gade}. 

We have also analysed various symmetry breaking terms. The simplest 
ones are those which spoil the particular symmetry 
Eq.(\ref{Summary:condition}) of the model at $E=0$, 
i.e. an on-site disorder or a same-sublattice regular hopping. 
These perturbations bring the symmetry of $U(R)$ 
down to Sp($2m$)/Sp($m$)$\times$ Sp($m$), 
as in the standard localization problem. However, by evaluating the 
anomalous dimensions of these terms, we can estimate the 
cross-over lengths above which the symmetry reduction is effective.     
While in $d=1,2$ these terms always lead to a localized behavior 
also at the band center, in $d=3$ the vicinity to the band center  
leads to an increase of the window in which delocalized states exist. 

Finally, if the impurity potential breaks time-reversal symmetry, the 
matrix field $U(R)$ is shown to belong to the coset space 
U$(2m)$, which indeed agrees with the analysis of Gade\cite{Gade}. 

The paper is organized as follows. In section \ref{The Model}, we 
introduce the Hamiltonian. In section \ref{Path Integral}, we derive 
the path-integral representation of the model, by using Grassmann variables 
within the replica trick method, and, in section \ref{Symmetries}, we study 
the symmetry properties of the action. In section \ref{Saddle Point}, we 
evaluate the saddle point of the action, while in sections 
\ref{Effective Action}, \ref{Longitudinal Fluctuations}, 
\ref{Effective NLSM} we derive the effective non-linear $\sigma$-model 
describing the long-wavelength fluctuations around the saddle point. 
The Renormalization Group analysis is presented in section 
\ref{Renormalization Group}, and the behavior in the presence of 
on-site disorder, of a same-sublattice regular hopping or with broken 
time reversal invariance is studied in sections \ref{On site disorder}, 
\ref{Same--Sublattice Regular Hopping}, and 
\ref{Time-reversal symmetry breaking}, respectively. Finally, 
section \ref{Conclusions} is devoted to a discussion of the results. 
We have also included several appendices containing more technical 
parts.

\section{The Model}
\label{The Model}

We consider a tight binding Hamiltonian on a bipartite lattice, of the form   
\be
H = \sum_{R\in A}\sum_{R'\in B} h_{RR'}\left(
\cc_R \ca_{R'} + \cc_{R'} \ca_{R}\right),
\label{Mod:Ham}
\ee
where $A$ and $B$ label the two sublattices and the hopping matrix 
elements $h_{RR'}$ are randomly distributed. 
We take a unit cell which  
includes two sites from different sublattices. 
In some cases, like the honeycomb lattice, this is indeed the primitive 
unit cell. In other cases, like the square lattice, it is not.

In this representation, the Hamiltonian can be written as 
\be
H = \sum_{RR'} h^{12}_{RR'}\left(\cc_{1R}\ca_{2R'} + 
                H.c.\right)
\label{Mod:Himp}
\ee
where $1$ and $2$ label now the two sites in the unit cell, 
while $R$ and $R'$ refer to the unit cells, and 
$h^{12}_{RR'} = h^{21}_{R'R}$. 
By introducing the two component operators  
\[
\ca_R =\left(
\begin{array}{c}
\ca_{1R}\\
\ca_{2R}\\
\end{array}
\right),
\]
we can also write the Hamiltonian as 
\bea
H &=& \sum_{R,R'} \cc_R\, H_{RR'}\,\ca_{R'} \nonumber \\
&=& 
\sum_{R,R'} 
\frac{1}{2}\left(h^{12}_{RR'} + h^{21}_{RR'}\right)
\cc_R \sigma_1 \ca_{R'}
+\frac{i}{2}\left(h^{12}_{RR'} - h^{21}_{RR'}\right)
\cc_R \sigma_2 \ca_{R'},
\label{Mod:FullHam}
\eea
where the $\sigma_i$'s ($i=1,2,3$) are the Pauli matrices. 
We notice that, quite generally, the Hamiltonian involves both  
$\sigma_1$ and $\sigma_2$, but neither $\sigma_3$ nor $\sigma_0$, 
so that it satisfies the 
conditions in Eq. (\ref{Summary:condition}).

We can write $h^{12}_{RR'}=t^{12}_{RR'} + \tau^{12}_{RR'}$, where  
$t^{12}_{RR'}$ are the average values, which represent the regular 
(translationally invariant) hopping 
matrix elements, 
while $\tau^{12}_{RR'}$ are random variables with zero average, which we 
assume to be gaussian distributed with width    
\[
\langle \left(\tau^{12}_{RR'}\right)^2 \rangle = 
u^2 \left(t^{12}_{RR'}\right)^2.
\]
The dimensionless parameter $u$ is a measure of the disorder 
strength in units of the regular hopping. In this way, the 
Hamiltonian is written as the sum of a regular part, $H^{(0)}$, 
plus a disordered part, $H_{imp}$.  

For the regular hopping, we define 
\[
t_{RR'} = \frac{1}{2}\left(t^{12}_{RR'} + t^{21}_{RR'}\right),\;\;
w_{RR'} = \frac{1}{2}\left(t^{12}_{RR'} - t^{21}_{RR'}\right),
\]
so that the non disorderd part, $H^{(0)}$, of the Hamiltonian is  
\be
H^{(0)} =  \sum_{R,R'} \cc_R\, H^{(0)}_{RR'}\,\ca_{R'} = 
\sum_{R,R'} \cc_R\left(t_{RR'}\sigma_1 +iw_{RR'}\sigma_2\right)\ca_{R'}.
\label{Mod:Ham2}
\ee
Since, for any lattice vector $R_0$, $t_{R\,R'} = t_{R+R_0\,R'+R_0}$, 
as well as $w_{R\,R'} = w_{R+R_0\,R'+R_0}$, and, moreover, 
$t_{RR'}=t_{R'R}$ while $w_{RR'}=-w_{R'R}$, 
the Fourier transforms satisfy
$t_k = t_k^*$ ($t_k$ real) and $w_k = -w_k^*$ ($w_k$ imaginary). 
In momentum space, the Hamiltonian matrix,   
$
H^{(0)}_k = t_k\sigma_1 +iw_k\sigma_2
$, 
is diagonalized by the unitary transformation $\ca_k = U_k \da_k$, with 
\be
U_k = {\rm e}^{-i\frac{\pi}{4}\sigma_2} {\rm e}^{i\frac{\theta_k}{2}\sigma_1},
\label{Mod:U}
\ee
where
\be
\tan\theta_k = \frac{iw_k}{t_k} = -
\frac{{\cal I}m \,t^{12}_k}{{\cal R}e\, t^{12}_k}.
\label{Mod:theta}
\ee
Indeed, 
$
U_k^\dagger H^{(0)}_k U_k = \epsilon_k \sigma_3
$, 
where $\epsilon_k^2 = t_k^2-w_k^2 = |t_k|^2 + |w_k|^2$.

\subsection{Current operator}
The commutator of the density $\cc_R\ca_R$ with the 
non disordered Hamiltonian 
(\ref{Mod:Ham2}) is
\[
\sum_{R'}
\cc_R\left(t_{RR'}\sigma_1 +iw_{RR'}\sigma_2\right)\ca_{R'}
-\cc_{R'}\left(t_{R'R}\sigma_1 +iw_{RR'}\sigma_2\right)\ca_{R},
\]
or, in Fourier space,  
\[
\sum_k \left(t_{k+q}-t_{k}\right) \cc_k\sigma_1\ca_{k+q}
+i\left(w_{k+q}-w_{k}\right) \cc_k\sigma_2\ca_{k+q}.
\]
Therefore, in the long wavelength limit, the current operator 
in the absence of disorder is 
\bea
\vec{J}^{~(0)}_q &&\simeq \sum_k \vec{\nabla}_k t_k \cc_{k}\sigma_1\ca_{k+q}
+i\vec{\nabla}_k w_k \cc_{k}\sigma_2\ca_{k+q}\nonumber \\
&&= \sum_k \vec{\nabla}_k\epsilon_k \,
\cc_k \vec{B}_k\cdot \vec{\sigma} \ca_k 
+\epsilon_k \vec{\nabla}\theta_k 
\cc_k \vec{B}_{\perp,k} \cdot \vec{\sigma} \ca_k 
\label{Current-c}\\
&&= \sum_k  \vec{\nabla}\epsilon_k\, \dc_{k}\sigma_3\da_{k+q}
+\epsilon_k\vec{\nabla}\theta_k \, \dc_{k}\sigma_2\da_{k+q},
\label{Mod:current}
\eea
the last being the expression in the basis which diagonalizes the 
Hamiltonian $H_0$. In (\ref{Current-c}), 
$\vec{\sigma}=(\sigma_1,\sigma_2,\sigma_3)$, and the vectors
\be 
\vec{B}_k = (\cos\theta_k,\sin\theta_k,0),\;\; 
\vec{B}_{\perp,k} = (-\sin\theta_k,\cos\theta_k,0),
\label{BkBperpk}
\ee 
describe intra and inter-band contributions to the current vertex.
Notice that the regular hopping Hamiltonian can be simply written as 
$H^{(0)} =\sum_k \epsilon_k \cc_k \,  \vec{B}_k\cdot \vec{\sigma}\, \ca_k$. 

Moreover, since also the impurity part of the whole 
Hamiltonian, (\ref{Mod:Himp}), does not 
commute with the density operator, in the disordered model the 
current operator acquires an additional term proportional 
to the random hopping matrix elements, which we discuss later.

\section{Path Integral}
\label{Path Integral}

The starting point of our analysis is a path-integral representation 
of the generating functional, in terms of Grassmann variables, 
following the work by Efetov, Larkin and Khmelnitskii\cite{EL&K}. 
To this end, we introduce, for each unit cell $R$, the Grassmann variables 
$c_{R;a,p,\alpha}$ and their {\sl complex conjugates} 
$\overline{c}_{R;a,p,\alpha}$, where $a=1,2$ is the sublattice index, 
$p=\pm$ is the index of the advanced (+) and retarded (-) components, 
and the index $\alpha$ runs over $m$ identical copies of the model, as in the 
usual replica trick method.    
In what follows, by convention, the Pauli matrices $\sigma$'s 
act in the two sublattice space, 
the $\tau$'s in the space of the Grassmann fields 
$c$ and $\overline{c}$, and the 
$s$'s in the $\pm$ space.

In order to treat on equal footing both the particle-hole  
and the particle-particle diffusive modes ({\sl diffusons} and 
{\sl cooperons}, respectively), as implied by time-reversal invariance, 
it is convenient to introduce the Nambu spinors $\Psi_R$ and 
$\overline{\Psi}_R$ defined through 
\[
\Psi_R =\frac{1}{\sqrt{2}} 
\left(
\begin{array}{c}
\overline{c}_R\\
c_R\\
\end{array}
\right),
\]
where $\overline{c}_R$ and $c_R$ are column vectors with components 
$\overline{c}_{R;a,p,\alpha}$ and $c_{R;a,p,\alpha}$, respectively, and  
\[
\overline{\Psi}_R = \left[ c\Psi\right]^t,
\]
where $c=-i\tau_2$ is the charge conjugation operator. 
The action in terms of the spinors is 
[see Eq.(\ref{Mod:FullHam})]
\bea
S &=& -\sum_{RR'} \overline{\Psi}_R 
\left( E + i\frac{\omega}{2}s_3 - H_{RR'} \right)
\Psi_{R'}\nonumber\\
&=& -\sum_{RR'} \overline{\Psi}_R 
\left( E + i\frac{\omega}{2}s_3 - H^{(0)}_{RR'} \right)
\Psi_{R'} 
+ \sum_{RR'} \overline{\Psi}_R\, H_{imp,RR'}\, 
\Psi_{R'}\nonumber\\
&=& S_0 + S_{imp},
\label{Path:Sinizio}
\eea
where $E\pm i\omega/2$ are the complex energies of the advanced/retarded 
components.  
\subsection{Disorder average}

Before taking the disorder average, we notice that, in the spinor notation,  
\[
\cc_{1R}\ca_{2R'} + H.c. \to 2\overline{\Psi}_{1R}\Psi_{2R'} 
= 2\overline{\Psi}_{2R'}\Psi_{1R}.
\]
Therefore, the impurity part of the action can be written as 
\[
S_{imp}= \sum_{R,R'} 2\tau^{12}_{RR'} \overline{\Psi}_{1R}\Psi_{2R'}.  
\]

The generating functional, within the replica method, is 
\be
{\cal Z} = \int {\cal D}\overline{\Psi}
{\cal D}\Psi {\cal D}\tau P[\tau] {\rm e}^{-S_0-S_{imp}},
\ee
where $P[\tau]$ is the gaussian probability distribution 
of the random bonds $\tau^{12}_{RR'}$. The  
average over disorder changes the impurity action into  
\bea
S_{imp} &=& -\sum_{R,R'} 2 u^2\left(t^{12}_{RR'}\right)^2 
\left(\overline{\Psi}_{1R}\Psi_{2R'}\right)^2\nonumber\\
&=&  -\sum_{R,R'} 2 u^2\left(t^{12}_{RR'}\right)^2 
\left(\overline{\Psi}_{1R}\Psi_{2R'}\right)
\left(\overline{\Psi}_{2R'}\Psi_{1R}\right).
\label{Path:Simp}
\eea
We define 
\[
W_{RR'}=2u^2\left(t^{12}_{RR'}\right)^2\in {\cal R}e,
\]
so that $W_q^*=W_{-q}$, and introduce 
\[
X^{\alpha\beta}_{1R} = \overline{\Psi}^\alpha_{1R}\Psi^\beta_{1R},
\]
where $\alpha$ is a multilabel for Nambu, advanced/retarded and 
replica components, and analogously 
$X^{\alpha\beta}_{2R}$, as well as their Fourier 
transforms. By these definitions,  
\be
S_{imp} = \frac{1}{V}\sum_q \sum_{\alpha,\beta} 
W_{-q} X^{\alpha\beta}_{1,q}
X^{\beta\alpha}_{2,-q}.
\label{Path:Simpbis}
\ee
This form, as compared to (\ref{Path:Simp}), has the advantage to 
allow a simple Hubbard--Stratonovich transformation.  
Notice that the use of Nambu spinors has the great advantage 
to involve just a single Fourier component of $W_{RR'}$.
If we write 
\[
W_q = \omega_q {\rm e}^{i\phi_q},
\]
where $\omega_q>0$ and $\phi_q=-\phi_{-q}$, and define 
\[
Y_{1q}={\rm e}^{-i\frac{\phi_q}{2}}X_{1q},\;\;
Y_{2q}={\rm e}^{i\frac{\phi_q}{2}}X_{2q},
\]
$S_{imp}$ takes the simple form 
\bea
S_{imp} &=& \frac{1}{V}\sum_q \omega_q Y^{\alpha\beta}_{1,q}
Y^{\beta\alpha}_{2,-q}
= \frac{1}{V}\sum_q \frac{\omega_q}{2} 
Tr\left[Y_{1,q}Y_{2,-q}+Y_{2,q}Y_{1,-q}\right]\nonumber\\
&=& \frac{1}{V}\sum_q \frac{\omega_q}{4} 
Tr\left[Y_{0,q}Y_{0,-q}-Y_{3,q}Y_{3,-q}\right],
\label{Path:Simp2}
\eea
where we have introduced $Y_0=Y_1+Y_2$ as well as $Y_3=Y_1-Y_2$. 
Moreover, our choice of the impurity potential, which does not break, 
on average, the spatial symmetries of the lattice, implies that 
$\phi_q=-\theta_q$, see Eq.(\ref{Mod:theta}).
    
We notice that, if a term is written as 
\[
\lambda \frac{A^2}{4} \sum_{\alpha\beta} X^{\alpha\beta}X^{\beta\alpha} 
= \lambda \frac{A^2}{4} Tr\left(X^2\right),
\]
where $X=X^\dagger$ and $\lambda=\pm 1$, one can always decouple it, by introducing an hermitean 
matrix $Q$, by the following Hubbard--Stratonovich transformation 
\be
\exp\left[\lambda \frac{A^2}{4} Tr\left(X^2\right)\right] = N 
\int {\cal D}Q\, \exp\left[ 
- A^{-2} Tr(Q^2) + \sqrt{\lambda} 
Tr\left(QX^t\right)\right],
\label{Path:Gaussian}
\ee
where the normalization factor 
$N^{-1}=\int {\cal D}Q\, \exp\left[ - A^{-2} Tr(Q^2)\right]$.
In the specific example, (\ref{Path:Simp2}) can be transformed into 
\bea
S_{imp} &=&
\frac{1}{V}\sum_q \frac{1}{\omega_q}Tr\left[ Q_{0q}Q_{0-q} + 
Q_{3q}Q_{3-q}\right]
- \frac{i}{V} \sum_q Tr\left[Q_{0q}Y^t_{0-q} + i Q_{3q}Y^t_{3-q}\right]
\nonumber\\
&=&
\frac{1}{V}\sum_q \frac{1}{\omega_q}Tr\left[ Q_{0q}Q_{0-q} + 
Q_{3q}Q_{3-q}\right]\nonumber\\
&-&\frac{i}{V} \sum_q Tr\left[Q_{0q}\left(\cos\frac{\phi_q}{2} X^t_{0-q} 
+ i\sin\frac{\phi_q}{2} X^t_{3-q}\right)\right]\nonumber\\
&-&\frac{i}{V} \sum_q iTr\left[Q_{3q}\left(\cos\frac{\phi_q}{2} X^t_{3-q} +
i\sin\frac{\phi_q}{2} X^t_{0-q}\right)\right].
\label{Path:Simp3}
\eea
If we define $Q_q = Q_{0q}\sigma_0 
+iQ_{3q}\sigma_3$, we obtain
\bea
S_{imp} &=&\frac{1}{V}  
\sum_q \frac{1}{2\omega_q}Tr\left[ Q_{q}^{\phantom{\dagger}}
Q_{q}^\dagger\right] -\frac{i}{V} \sum_{p,q} 
\overline{\Psi}_{p} Q_{-q}{\rm e}^{-\frac{i}{2}\phi_q\sigma_3}
\Psi_{p+q}\label{Path:Simp3a}\\
&=&  \frac{1}{V}
\sum_q \frac{1}{2\omega_q}Tr\left[ Q_{q}^{\phantom{\dagger}}
Q_{q}^\dagger\right] - i\sum_{R} 
\overline{\Psi}_{R} Q_{R}
\Psi_{R}
+\frac{i}{V} \sum_{p,q} 
\overline{\Psi}_{p} 
\left(1-{\rm e}^{-\frac{i}{2}\phi_q\sigma_3}\right)Q_{-q}
\Psi_{p+q},
\label{Path:Simp3b}
\eea
where the last term vanishes at $q=0$.
Notice that the two-sublattice symmetry properties of 
the Hamiltonian, see Eq. (\ref{Summary:condition}), are reflected in the 
particular form of the $Q$-matrix, which only contains the Pauli matrices 
$\sigma_0$ and $\sigma_3$. In particular, since the tensor  
$Q(R)\sim \Psi(R)\overline{\Psi}(R)$, $\sigma_0$ selects the uniform 
component while $\sigma_3$ the staggered component of the product of the 
two Grassmann fields.  
Notice that the electron-$Q$ coupling in the $\Psi_R$ basis can be 
written as a local coupling but for the last term in (\ref{Path:Simp3b}). 
To simplify the notation of this term we find it useful to 
define the operator $\hat{L}$ through  
\be
\hat{L}Q(R) = \frac{1}{V}\sum_q {\rm e}^{iqR}
\left(1-{\rm e}^{\frac{i}{2}\phi_q\sigma_3}\right)Q_q.
\label{Def:L}
\ee

If we now transform the 
spinors in Eq.(\ref{Path:Simp3a}) to the diagonal basis, the coupling 
to the $Q_{q}$ 
matrix transforms as  
$U^\dagger_{p+q} Q_{q} {\rm e}^{\frac{i}{2}\phi_q\sigma_3}
U^{\phantom{\dagger}}_{p}$. 
The simplest consequence is that, in the diagonal basis, 
$Q_{k_1,k_2}=Q_{0,k_1,k_2}\sigma_0+iQ_{1,k_1,k_2}\sigma_1$ depends 
on two wavevectors.       
However, in the case of cubic lattices, see Eq.(\ref{thetasquare}), 
\ben
&&U^\dagger_{p+q} Q_{q}{\rm e}^{\frac{i}{2}\phi_q\sigma_3}
 U^{\phantom{\dagger}}_{p+q} = 
{\rm e}^{-\frac{i}{2}\theta_{p+q}\sigma_1}\left(
Q_{0,q} - iQ_{3,q}\sigma_1 \right)
{\rm e}^{\frac{i}{2}\theta_{p}\sigma_1} {\rm e}^{-\frac{i}{2}\phi_q\sigma_1}\\
&&= {\rm e}^{-\frac{i}{2}(\theta_{q}+\phi_q)\sigma_1}
\left(
Q_{0,q} - iQ_{3,q}\sigma_1 \right)= Q_{0,q} - iQ_{3,q}\sigma_1,
\een
since $\phi_q=-\theta_q$.
Therefore, for cubic lattices, we can also write 
\be
S_{imp}=\frac{1}{V}
\sum_q \frac{1}{2\omega_q}Tr\left[ Q_{q}^{\phantom{\dagger}}
Q_{q}^\dagger\right]
- i\sum_R \overline{\Phi}_R Q(R) \Phi_R,
\label{Simpsquare}
\ee
where now $\Phi_R$ 
is the Grassmann field of the $\da_R$ operators, and the 
matrix $Q(R) = Q_0(R)\sigma_0 
+ iQ_1(R)\sigma_1$, with $Q_1(R)=-Q_3(R)$ defined above.  

Finally, we notice that, in the case of the honeycomb lattice, 
at the wavevector $q_*$, connecting the two Dirac cones of the 
non-disordered dispersion band, $\omega_{q_*}=0$. This observation 
will turn useful when discussing the long-wavelength behavior of the model.   

\section{Symmetries}
\label{Symmetries}

The action (\ref{Path:Sinizio}), at $E=\omega=0$, i.e. at the band center 
with zero complex frequency, is invariant under a 
transformation $\Psi_R \to T\,\Psi_R$ if 
\[
c\, T^t\, c^t\, H_{RR'}\, T = H_{RR'}.
\]
Since the random matrix elements $H_{RR'}$ involve both Pauli matrices 
$\sigma_1$ and $\sigma_2$, $T$ has to satisfy at the same time   
$
c\, T^t\, c^t\, \sigma_1 \, T = \sigma_1$ and 
$
c\, T^t\, c^t\, \sigma_2 \, T = \sigma_2$. This implies that 
\bea
c\, T^t\, c^t\ &=& \sigma_1 \, T^{-1}\, \sigma_1 \label{Sym:uno}\\
\sigma_1 \, T^{-1}\, \sigma_1 &=& 
\sigma_2 \, T^{-1}\, \sigma_2.\label{Sym:due}
\eea
The condition (\ref{Sym:due}) can be fulfilled only by a transformation 
$T=T_0\sigma_0 + T_3\sigma_3$ 

Under such a transformation 
\[
Q \to cT^t c^t Q T = \sigma_1 T^{-1}\sigma_1 Q T \equiv 
  \sigma_2 T^{-1}\sigma_2 Q T
\]
Since $\sigma_1 Q \sigma_1 = Q^\dagger$, then 
\[
T^{-1}\sigma_1 Q T\sigma_1 = 
T^\dagger Q^\dagger \sigma_1 
\left(T^{-1}\right)^\dagger \sigma_1 =  
T^\dagger \sigma_1 Q \left(T^{-1}\right)^\dagger 
\sigma_1.
\]
Hence, the transformation is also unitary, $T^\dagger=T^{-1}$. 
Moreover, such a transformation   
leaves the $Q$-manifold invariant, which implies that our Hubbard-Stratonovich 
decoupling scheme, which makes use of $Q=Q_0\sigma_0 + iQ_3\sigma_3$, 
is exhaustive.    

The unitary transformation, $T$, can be written as  
\be
T = \exp\left[ \frac{W_0}{2}\sigma_0 + \frac{W_3}{2}\sigma_3\right],
\label{TPsi}
\ee
where 
\[
W_0^\dagger = -W_0,\;\;\; W_3^\dagger = -W_3.
\]
In addition, we must impose the charge conjugacy invariance, which,  
through Eq.(\ref{Sym:uno}), implies that  
\[
\begin{array}{ccccc}
cW_0^tc^t & = & -W_0 & = & W_0^\dagger,\\
cW_3^tc^t & = & W_3  & = & -W_3^\dagger.\\
\end{array}
\]
The number of independent parameters turns out to be $16m^2$, which 
suggests that $T$ is related to a unitary group, specificaly U$(4m)$, as 
argued by Gade and Wegner\cite{W&G}. In fact, we can 
alternatively write 
\be
T = \left(
\begin{array}{cc}
{\rm e}^{\frac{W_0+W_3}{2}} & 0 \\
0 &  {\rm e}^{\frac{W_0-W_3}{2}} \\
\end{array}
\right)
\equiv 
\left(
\begin{array}{cc}
U & 0 \\
0 &  c\,U^\dagger\, c^t \\
\end{array}
\right),
\label{T-unitary}
\ee
where $U$ is indeed a unitary transformation belonging to  
U$(4m)$. 
The invariance of (\ref{Path:Sinizio}) at finite frequency, 
$\omega\not = 0$, implies the additional condition  
$c T^t c^t s_3 T = s_3$, which reduces the number of independent parameters 
to $8m^2 + 2m$, lowering the symmetry of $U$ 
down to Sp$(2m)$.   

If $E\not =0$, $T$ has to satisfy also 
\[
c T^t c^t T = \sigma_1 T^{-1} \sigma_1 T = 1.
\]
This implies that, at finite energy, i.e. away from the 
band center, $T$ does not contain anymore 
a $\sigma_3$-component. Indeed $E$ lowers the symmetry of $U$ down 
to Sp$(2m)$, which is further reduced to 
${\rm Sp}(m)\times {\rm Sp}(m)$ by a finite 
frequency, as in the standard situation\cite{EL&K}.  

\section{Saddle Point}
\label{Saddle Point}

The full action  
\bea
S &=& -\sum_{k,q}
\overline{\Psi}_k \left(E\delta_{q0} + i\frac{\omega}{2} s_3\delta_{q0} 
- H^{(0)}_{k}\delta_{q0} +
\frac{i}{V}Q_{-q} {\rm e}^{-\frac{i}{2}\phi_q\sigma_3}\right)
\Psi_{k+q}\nonumber\\
&+&\frac{1}{V}\sum_q \frac{1}{2\omega_q}Tr\left[ Q_{q}^{\phantom{\dagger}}
Q_{q}^\dagger\right],
\label{FullAction}
\eea
by integrating over the Nambu spinors, transforms into  
\be
S[Q] = \frac{1}{V}\sum_q \frac{1}{2\omega_q} 
Tr\left[Q_{q}Q_{q}^\dagger\right] -
\frac{1}{2}Tr\ln\left[E + i\frac{\omega}{2}s_3 - H^{(0)} 
+i Q - i\hat{L}Q\right].
\ee
The saddle point equation for homogeneous solutions at $E=0$ and   
$\omega=0^+$ is 
\[
Q = i\frac{\omega_0}{4}\int \frac{d^2 k}{4\pi^2} 
\left(i 0^+ s_3 - \epsilon_k + iQ\right)^{-1} +
\left(i 0^+ s_3 + \epsilon_k + iQ\right)^{-1} ,
\]
where $\omega_0 = \sum_{R-R'} 2u^2(t^{12}_{R-R'})^2$.
The general solution is 
\be
Q_{sp} = \frac{\pi}{4} \omega_0 \rho(0) s_3 \equiv \Sigma s_3, 
\label{Q:saddle_point}
\ee
with $\rho(0)$ being the density of states at $E=0$. 
In order to distinguish transverse from longitudinal modes, it is 
convenient to parametrize the $Q$-matrix in the following way  
\be
Q(R)_P = \sigma_1 T(R)^{-1} \sigma_1 \left[Q_{sp}+P(R)\right] T(R)
\equiv Q(R) + \sigma_1 T(R)^{-1} \sigma_1 P(R)T(R).
\label{QQ}
\ee
Here $P(R)$ describes the longitudinal modes, which we discuss more in detail 
in section \ref{Longitudinal Fluctuations}, and $T$ the transverse modes. 
Namely, $T$ has the form given in Eq.(\ref{T-unitary}), 
\be
T(R) = \exp \left(\frac{W(R)}{2}\right) = 
\exp\left(\frac{W_0(R)}{2}\sigma_0 + 
\frac{W_3(R)}{2}\sigma_3\right),
\label{T:matrix}
\ee
with 
$\exp[(W_0+W_3)/2]$ belonging now to the coset space 
${\rm U}(4m)/{\rm Sp}(2m)$. 
This amounts to impose that 
\[
\left\{W_0,s\right\}=0,\;\;\; \left[W_3,s\right]=0,
\]
by which it derives that 
\be
Q(R) = \sigma_1 T(R)^{-1} \sigma_1 Q_{sp}T(R) 
= Q_{sp} {\rm e}^{W_0(R)\sigma_0+W_3(R)\sigma_3}.
\label{Q:matrix}
\ee

In the $\pm$ space, we can write 
\be
W_0 = 
\left(
\begin{array}{cc}
0 & B\\
-B^\dagger & 0 \\
\end{array}
\right),\;\;
W_3 = 
\left(
\begin{array}{cc}
iA & 0\\
0& iC \\
\end{array}
\right),
\label{Def:W}
\ee
where $A^\dagger=A$, $C^\dagger=C$, and additionally, since 
$cW^tc^t=\sigma_1 W^\dagger\sigma_1=-\sigma_1W\sigma_1$, then   
$cA^tc^t=A$, $cC^tc^t=C$ and $cB^tc^t=B^\dagger$. By writing 
\bea
A&=&A_0\tau_0 + i\left(A_1\tau_1+A_2\tau_2+A_3\tau_3\right),\label{A:def}\\
B&=&B_0\tau_0 + i\left(B_1\tau_1+B_2\tau_2+B_3\tau_3\right),\label{B:def}\\
C&=&C_0\tau_0 + i\left(C_1\tau_1+C_2\tau_2+C_3\tau_3\right),\label{C:def}
\eea
we find that the above conditions imply that, for $i=0,\dots,3$, 
\be
B_i,A_i,C_i\in {\cal R}e,
\label{ABC:condition_real}
\ee
and 
\be
A_0=A_0^t,\;\; C_0=C_0^t,
\label{A0C0:condition}
\ee
while, for $j=1,2,3$,
\be 
A_j=-A_j^t,\;\; C_j=-C_j^t.
\label{AiCi:condition}
\ee 

\section{Effective Action}
\label{Effective Action}

In this section, we derive the effective field theory describing the 
long wavelength transverse fluctuations of $Q(R)$ around the saddle 
point. In the case of honeycomb lattices, 
we should worry about the momentum component of $Q$ which couples the two 
Dirac cones. However, one can see that the free action of $Q$ diverges 
at this wavevector, so that we are allowed to ignore the fluctuations 
around this momentum.  

\subsection{Integration over the Grassmann fields}

As we said, by integrating (\ref{FullAction}) over the Grassmann 
variables, we obtain the following action of $Q$:   
\be
-S[Q] = -\frac{1}{V}\sum_q \frac{1}{2\omega_q} 
Tr\left[Q_{q}Q_{q}^\dagger\right] +
\frac{1}{2}Tr\ln\left[E + i\frac{\omega}{2}s_3 - H^{(0)} 
+i Q - i\hat{L}Q\right].
\ee
We start by neglecting the longitudinal fluctuations. Then, since 
$Q=\tilde{T}^\dagger Q_{sp} T$, where we define 
$\tilde{T}=\sigma_1T\sigma_1\equiv \sigma_2T\sigma_2$, 
we can rewrite the second term of $S[Q]$ as
\bea
&&\frac{1}{2} Tr\ln\left(E\tilde{T}T^\dagger 
+i\frac{\omega}{2}\tilde{T}s_3T^\dagger 
- \tilde{T}H^{(0)}T^\dagger + iQ_{sp} - V
 \right)
\label{intermedio},
\eea
where we define 
\[
V = i\tilde{T}\hat{L}QT^\dagger.
\]

Since $H^{(0)}_{RR'}$ involves either $\sigma_1$ and $\sigma_2$, 
while $T$ involves $\sigma_0$ and $\sigma_3$, then 
\begin{eqnarray*}
H^{(0)}_{RR'}T(R')^\dagger 
= \tilde{T}(R')^\dagger H^{(0)}_{RR'}
&=& \tilde{T}(R)^\dagger H^{(0)}_{RR'} 
+\left(\tilde{T}(R')^\dagger-\tilde{T}(R)^\dagger\right)H^{(0)}_{RR'}\\
&=& \tilde{T}(R)^\dagger H^{(0)}_{RR'} 
-\vec{\nabla}\tilde{T}(R)^\dagger\cdot 
\left(\vec{R}-\vec{R'}\right) H^{(0)}_{RR'}\\
&+&\frac{1}{2}\partial_{ij}\tilde{T}(R)^\dagger 
\left(R_i-R'_i\right)\left(R_j-R'_j\right)H^{(0)}_{RR'}.
\end{eqnarray*}
Therefore the term $\tilde{T}H^{(0)}T^\dagger$ which appears in 
(\ref{intermedio}) can be written at long 
wavelengths as 
\bea
\tilde{T}(R) H^{(0)}_{RR'} T(R')^\dagger 
&=& H^{(0)}_{RR'} -i\tilde{T}(R)\vec{\nabla} \tilde{T}^{-1}(R)
\cdot \vec{J}^{~(0)}(R')_{RR'} \nonumber\\
&+& \frac{1}{2}\tilde{T}(R)\partial_{ij}\tilde{T}(R)^{-1} 
\left(R_i-R'_i\right)\left(R_j-R'_j\right)H^{(0)}_{RR'}\nonumber\\
&&\equiv H^{(0)}_{RR'} + U_{RR'},
\label{Act:U}
\eea
where we used the fact that the long-wavelength part of the current operator 
in real space is, see (\ref{Ward:current}), 
\[
\vec{J}(R') = -i\sum_R \left(\vec{R}-\vec{R'}\right)
\cc_{R}H_{RR'}\ca_{R'}.
\]
Notice that in (\ref{Act:U}) only the current vertex which derives from the 
regular hopping appears, which is not the full current operator. 

Moreover, a further current-like coupling will arise from the expansion in $V$ 
(see below).
To this end, in the long-wavelength limit, the operator $\hat{L}$, see 
(\ref{Def:L}), can be approximately written as 
\be
\hat{L}Q(R) \simeq -\vec{\beta}\cdot\vec{\nabla} Q(R)\sigma_3 
-\frac{1}{2}\left(\vec{\beta}\cdot\vec{\nabla}\right)^2 Q(R),
\label{EXP:L}
\ee
where $\vec{\beta} = \vec{\nabla}\phi(q=0)/2$. 

Having defined $U$, (\ref{intermedio}) can be written as 
\begin{eqnarray}
&&\frac{1}{2} Tr\ln\left(E\tilde{T}T^\dagger 
+i\frac{\omega}{2}\tilde{T}s_3T^\dagger 
- U - V - H^{(0)} + iQ_{sp}\right)\nonumber \\
&=&  -\frac{1}{2}Tr\ln G + 
\frac{1}{2} Tr\ln\left(1 + G\, E\tilde{T}T^\dagger 
+ G\,i\frac{\omega}{2}\tilde{T}s_3T^\dagger 
-G\, U - G\,V \right),
\label{43bis}
\eea
where $G=(-H^{(0)}+iQ_{sp})^{-1}$ is the Green's function in the absence of 
transverse fluctuations. 

The effective field theory is then derived by expanding 
$S[Q]$ up to second order in $U$ and $V$, and first order in $E$ and 
$\omega$. In this way we get the following terms.

\subsection{Expansion in the $Q$ free action}

The free part of the action
\[
S_0[Q] = \frac{1}{V}\sum_q \frac{1}{2\omega_q} 
Tr\left[Q_{q}Q_{q}^\dagger \right],
\]
can be expanded at small q. Since $\omega_q=\omega_{-q}$, then 
\[
\omega_q \simeq \omega_0 (1 - \gamma q^2),
\]
leading to 
\bea
S_0[Q] &\simeq& \frac{1}{2\omega_0} \int dR\,
Tr\left[ Q(R)Q(R)^\dagger\right] 
+\frac{\gamma}{2V \omega_0} \sum_q q^2 Tr\left[Q_q Q_q^\dagger\right]
\nonumber\\
&=& \frac{1}{2\omega_0} \int dR\,
Tr\left[ Q_{sp}^2\right] 
+\frac{\gamma}{2\omega_0} 
\int dR\, Tr\left[\vec{\nabla}Q(R) 
\cdot \vec{\nabla}Q(R)^\dagger\right].
\label{SfreeQ}
\eea
The second term is a contribution to the current-current 
correlation function of the part of the current vertices proportional 
to the random hopping.    

\subsection{Expansion in $E$}

Expansion of (\ref{43bis}) in $E$ gives
\be
\frac{E}{2}Tr\left(G\,\tilde{T}T^\dagger\right) 
= -i \frac{E}{\omega_0}Tr\left(Q_{sp}\tilde{T}T^\dagger\right) = 
-i \frac{E}{\omega_0}Tr Q.
\label{Exp:E}
\ee
 
\subsection{Expansion in $\omega$}
Expansion in $\omega$ gives
\be
i\frac{\omega}{4}Tr\left(G\,\tilde{T}\hat{s}T^\dagger\right)
=\frac{\omega}{2\omega_0}Tr\left(s_3 Q\right).
\label{Exp:w} 
\ee

\subsection{Expansion in $U$}

The second order expansion in $U$ contains the following terms:
\be
\label{Act:first}
-\frac{1}{2}Tr\left(G\,U\right),
\ee
and
\be
-\frac{1}{4}Tr\left(G\, U\, G\, U\right),
\label{Act:second}
\ee
Taking in (\ref{Act:first}), the component of $U$ containing second 
derivatives, we get
\[
-\frac{1}{4}Tr\left\{\tilde{T}(R)\partial_{ij}\tilde{T}(R)^{-1} 
\left(R_i-R'_i\right)\left(R_j-R'_j\right)H^{(0)}_{RR'}G(R',R)\right\}.
\]
By means of the Ward identity (\ref{Ward3}), the above 
expression turns out to be  
\be
-\frac{\chi^{++}_{ij}}{8}
Tr\left\{\tilde{T}(R)\partial_{ij}\tilde{T}(R)^{-1} \right\},
\ee
which, integrating by part, is also equal to 
\bea
&-&\frac{\chi^{++}_{ij}}{8}
Tr\left\{\tilde{T}(R)\partial_{i}\tilde{T}(R)^{-1} 
\tilde{T}(R)\partial_{j}\tilde{T}(R)^{-1}\right\}\nonumber\\
&=&-\frac{1}{16}\chi^{++}_{ij}
Tr\left[ D_i D_j -
D_is_3\sigma_1 D_j s_3\sigma_1\right]\label{Act:diama}
\\
&-&\frac{1}{16}\chi^{++}_{ij}
Tr\left[ D_i D_j + 
D_is_3\sigma_1 D_j s_3\sigma_1\right].
\label{Act:diama1}
\eea
Here we have introduced a matrix $\vec{D}(R)$ with the $i$-th component 
\be
D_i(R) 
= D_{0,i}(R)\sigma_0 + D_{3,i}(R)\sigma_3 \equiv 
\tilde{T}(R)\partial_i \tilde{T}(R)^{-1}.
\label{DR}
\ee
The part of (\ref{Act:first}) which contains first derivatives gives rise 
to a boundary term 
\be
\left[ \frac{1}{V}\sum_k \vec{\nabla}\theta_k 
\frac{\epsilon_k^2}{\epsilon_k^2 + \Sigma^2}\right] 
\int dR\, Tr\left[\vec{\nabla}W(R)\sigma_3\right],
\label{Boundary}
\ee
where $\Sigma$ has been defined by Eq.(\ref{Q:saddle_point}), which we 
discard by taking appropriate boundary conditions.

Let us now analyse the term (\ref{Act:second}), where we have to keep 
of $U$ only the part containing first derivatives. 
By making use of (\ref{Act:U}), 
this term is, in momentum space,
\bea
&&\frac{1}{4}\sum_{kq}Tr\left\{
\left[\tilde{T}\vec{\nabla}\tilde{T}^{-1}\right]_q 
\cdot \vec{J}^{~(0)}_{k+q} G(k+q)\,
\left[\tilde{T}\vec{\nabla}\tilde{T}^{-1}\right]_{-q} 
\cdot \vec{J}^{~(0)}_k G(k)\right\}\nonumber\\
&&\simeq\frac{1}{4}\sum_{kq}Tr\left\{
\left[\tilde{T}\vec{\nabla}\tilde{T}^{-1}\right]_q 
\cdot \vec{J}^{~(0)}_{k} G(k)\,
\left[\tilde{T}\vec{\nabla}\tilde{T}^{-1}\right]_{-q} 
\cdot \vec{J}^{~(0)}_k G(k)\right\}\nonumber\\
&&=\frac{1}{4}\sum_{k}\sum_R \, Tr\left\{
\vec{D}(R)
\cdot \vec{J}^{~(0)}_{k} G(k)\,
\vec{D}(R)
\cdot \vec{J}^{~(0)}_k G(k)\right\},\nonumber\\
\label{Act2}
\eea
valid for small $q$. Here the matrix 
$
\vec{J}^{~(0)}_k = 
\vec{\nabla}_k t_k\sigma_1 + i\vec{\nabla}_k w_k \sigma_2
$.
The non vanishing terms in (\ref{Act2})   
have both $\vec{D}$'s either $\vec{D}_0\sigma_0$ or 
$\vec{D}_3\sigma_3$ [see (\ref{DR})]. 

In the diagonal basis, upon defining, as we did in 
Eq.(\ref{Q:saddle_point}), $Q_{sp}=\Sigma\,s_3$, 
with $\Sigma = \pi\omega_0\rho(0)/4$, the Green's function is 
\bea
{\cal G}(k) 
&=& \frac{1}{-\epsilon_k\sigma_3 + i\frac{\pi}{4}\omega_0\rho(0)s_3} 
\nonumber\\
&=& 
-i\frac{\Sigma}{\epsilon_k^2+\Sigma^2}\sigma_0s_3 
-\frac{\epsilon_k}{\epsilon_k^2+\Sigma^2}\sigma_3s_0\nonumber\\
&\equiv& G_0(k)\sigma_0s_3 + G_3(k)\sigma_3s_0. 
\label{Greenfunctiond}
\eea
Going back to the original basis, 
\be
G(k) = U_k {\cal G}(k) U_k^\dagger 
= G_0(k)\sigma_0s_3 +G_3(k) \vec{B}_k\cdot\vec{\sigma} s_0, 
\label{Greenfunction}
\ee
where $\vec{B}_k$ has been defined in (\ref{BkBperpk}).
Therefore, 
\be
\sigma_3 G(k) \sigma_3 = G_0(k)\sigma_0s_3 - G_3(k) 
\vec{B}_k\cdot\vec{\sigma} s_0 = - s_1 G(k) s_1.
\label{G:Identity}
\ee
By means of (\ref{G:Identity}), we find that 
\be
\sigma_3 \vec{J}^{~(0)}_k G(k)^+\sigma_3 = 
-\vec{J}^{~(0)}_k \sigma_3G(k)^+\sigma_3 
= \vec{J}^{~(0)}_k G(k)^-,
\ee
from which it derives that (\ref{Act2}) can be written as the sum of 
two different terms
\bea
&&\frac{1}{16}\chi^{+-}_{ij}Tr\left[
D_iD_j - D_i s_3\sigma_1D_j s_3\sigma_1\right]\label{Act21}\\
&&+\frac{1}{16}\chi^{++}_{ij}Tr\left[
D_iD_j + D_is_3\sigma_1D_j s_3\sigma_1\right].\label{Act22}
\eea
By summing (\ref{Act21}), (\ref{Act22}), (\ref{Act:diama}) 
and (\ref{Act:diama1}), we get 
\[
\frac{1}{16}\left(\chi^{+-}_{ij}-\chi^{++}_{ij}\right)Tr\left[
D_iD_j - D_is_3\sigma_1 D_j s_3\sigma_1\right],
\]
which is equal to 
\be
-\frac{2\pi}{32\Sigma^2}\sigma^{(0)}_{ij}
\int dR\, Tr\left(\partial_i Q(R) \partial_j Q(R)^\dagger\right),
\label{Act:stiff}
\ee
where 
\[
\sigma^{(0)}_{ij} = \frac{1}{2\pi}\left(\chi^{+-}_{ij}-\chi^{++}_{ij}\right)
\]
is the Kubo conductivity with the current vertices which involve only the 
regular hopping [cf. Eq.(\ref{B4}]. 
\subsection{Expansion in $V$}
The expansion in $V$ up to second order, gives two terms 
\be
-\frac{1}{2}Tr\left(G\,V\right),
\label{V:first}
\ee
and 
\be
-\frac{1}{4}Tr\left(G\,V\,G\,V\right),
\label{V:second}
\ee
In addition, we must also consider the mixed term
\be
-\frac{1}{2}Tr\left(G\,U\,G\,V\right).
\label{UVmix}
\ee
The first order term (\ref{V:first}) is 
\bea
&&-\frac{1}{2}Tr\left(G\,V\right) = 
\frac{i}{\omega_0}Tr\left(Q_{sp}V\right) \nonumber\\
&&=\frac{1}{\omega_0}\int dR\, 
Tr\left(Q(R)^\dagger \vec{\beta}\cdot\vec{\nabla}Q(R)\sigma_3\right)
\label{boundary:V}\\
&&-\frac{1}{2\omega_0}\int dR\, 
Tr\left(\vec{\beta}\cdot\vec{\nabla}Q(R)^\dagger 
\vec{\beta}\cdot\vec{\nabla}Q(R)\right).
\label{V:dQdQ}
\eea
The first term (\ref{boundary:V}) is another boundary term, which we 
neglect.
The second order term, Eq.(\ref{V:second}), gives 
\be
-\frac{1}{4V}\sum_k \left(G_0(k)^2 + G_3(k)^2\right)
\int dR\, Tr\left[\vec{\beta}\cdot\vec{\nabla}Q(R)^\dagger 
\vec{\beta}\cdot\vec{\nabla}Q(R)\right].
\label{VII:dQdQ}
\ee
Notice that the saddle point equation implies that 
\[
\frac{1}{V}\sum_k G_3(k)^2-G_0(k)^2 = \frac{2}{\omega_0}.
\]
By using the above equation in (\ref{V:dQdQ}), we find that 
(\ref{VII:dQdQ}) plus (\ref{V:dQdQ}) give 
\bea
&&-\frac{1}{2V}\sum_k G_3(k)^2
\int dR\, Tr\left[\vec{\beta}\cdot\vec{\nabla}Q(R)^\dagger 
\vec{\beta}\cdot\vec{\nabla}Q(R)\right]\nonumber\\
&& = -\frac{1}{2V}\sum_k \frac{\epsilon_k^2}{\epsilon_k^2+\Sigma^2}
\int dR\, Tr\left[\vec{\beta}\cdot\vec{\nabla}Q(R)^\dagger 
\vec{\beta}\cdot\vec{\nabla}Q(R)\right]
\label{Vend:dQdQ}
\eea

The contribution of the mixing term, (\ref{UVmix}), can be 
evaluated in a similar way, giving 
\be
\frac{1}{2V}\sum_k \frac{\epsilon_k^2}{\epsilon_k^2+\Sigma^2}
\int dR\, Tr\left[\vec{\nabla}\theta_k \cdot\vec{\nabla}Q(R)^\dagger 
\vec{\beta}\cdot\vec{\nabla}Q(R)\right].
\label{UVmixend}
\ee 
We notice that, since $\phi_q=-\theta_q$, 
then (\ref{Vend:dQdQ}) and (\ref{UVmixend}) can be 
included in (\ref{Act:stiff}) if the following redefinition of the  
current vertex is assumed:
\be
\vec{J}_k^{~(0)}= 
\vec{\nabla}\epsilon_k\, \vec{B}_k\cdot\vec{\sigma}
+\epsilon_k
\left(\vec{\nabla}\theta_k-\vec{\nabla}\theta_0\right)  
\vec{B}_{\perp,k}\cdot\vec{\sigma}.
\ee
Therefore the Kubo conductivity which appears in (\ref{Act:stiff}) 
has to be calculated with the above expression of the regular current 
vertex. This has notable consequences. First of all, for cubic 
lattices, the interband contribution vanishes, hence the 
Kubo conductivity coincides with the true one, since the 
enlargment of the unit cell was artificial. This is compatible with 
(\ref{Simpsquare}), where we showed that the impurity action 
is local in the basis which diagonalizes the regular hopping, implying 
that the regular current vertex contains only the intraband operator.

To conclude, the effective action so far derived is therefore 
\bea
S[Q] &=&
\frac{1}{2\omega_0} \int dR\,
Tr\left[ Q(R)Q(R)^\dagger\right] \nonumber\\
&& + 
\frac{\gamma}{2V\omega_0} \sum_q q^2 Tr\left[Q_q Q_q^\dagger\right]
\nonumber\\
&+&\frac{2\pi}{32\Sigma^2}\sigma^{(0)}_{ij}
\int dR\, Tr\left(\partial_i Q(R) \partial_j Q(R)^\dagger\right)
\nonumber\\
&+& \int dR\, i \frac{E}{\omega_0}Tr\left(Q(R)\right)
- \frac{\omega}{2\omega_0}Tr\left(s_3 Q(R)\right).
\label{SQ1}
\eea 


\section{Longitudinal Fluctuations}
\label{Longitudinal Fluctuations}

The full expression of the $Q$-matrix we must indeed consider is 
the one given by (\ref{QQ}), 
\bea
Q(R)_P &=& \tilde{T}(R)^{-1} \left[Q_{sp} + P(R)\right] T(R) \equiv 
Q(R) + \tilde{T}(R)^{-1} P(R) T(R)\nonumber\\
&\equiv& Q(R) + S(R),
\label{QPR}
\eea
where $T(R)$ involves transverse fluctuations and 
\be
P(R) = \left(P_{00}s_0 + P_{03}s_3\right)\sigma_0
+ i\left(P_{31}s_1 + P_{32}s_2\right)\sigma_3,
\label{PR}
\ee
being all $P$'s hermitean. Charge conjugation implies that 
$
cP^tc^t = P
$.
The field $P(R)$ takes into account longitudinal fluctuations which 
are massive. Let us define $\Gamma(R-R')$ the Fourier transform of 
$\omega_q^{-1}$. Then, the free action of the $Q_P$ field is 
\bea
S[Q_P] &=& \frac{1}{2}\int dR\,dR'\, \Gamma(R-R') Tr\left[Q_P(R)
Q_P(R')^\dagger\right]
\nonumber\\
&=& \frac{1}{2\omega_0}\int dR\, Tr\left[Q_P(R)Q_P(R)^\dagger\right] 
\label{SQ_Ploc}\\
&&- \frac{1}{4}\int dR\,dR'\, \Gamma(R-R')\, 
Tr\left[\left(Q_P(R)^{\phantom{\dagger}}- 
Q_P(R')^{\phantom{\dagger}}\right)
\left(Q_P(R)^\dagger-Q_P(R')^\dagger\right)\right]. 
\label{SQ_P}
\eea
Since $QQ^\dagger=Q_{sp}^2$, (\ref{SQ_Ploc}) gives
\[
\frac{1}{2\omega_0}\int dR Tr\left[P(R)P(R)^\dagger + 2Q_{sp}P(R)
+ Q_{sp}^2 \right].
\] 
The second term cancels with the first order expansion of $Tr\ln G_P$, 
since $Q_{sp}$ is the saddle point solution. What is left, i.e.
\be
\frac{1}{2\omega_0}\int dR\, Tr\left[P(R)P(R)^\dagger \right],
\label{Along-loc}
\ee
is actually the mass term of the longitudinal modes, since the second order 
expansion in $P$ of $Tr\ln G_P$ is zero. The other term, (\ref{SQ_P}), 
can be analysed within a gradient expansion of 
$Q_P(R)-Q_P(R') = \vec{\nabla}Q_P(R) \cdot \left(\vec{R}-\vec{R'}\right) 
+ \dots$. The details of the calculations are given  
in Appendix \ref{Appendix:Longitudinal}, so that, in this section, we just 
present the final results. 

The free action of the longitudinal fields is found to be 
\be
S_0[P] = \frac{1}{V} \sum_q \frac{1}{2\omega_q} Tr(P_qP_{q}^\dagger).
\label{Along}
\ee
Here, we neglect the 
the contribution of the invariant measure, which, in the zero 
replica limit, gives rise to fluctuations smaller by a factor  
$u^{2}$ than Eq.(\ref{Along}).
The integration over $P$ with the above action has several important 
consequences for the action of the transverse modes 
(see Appendix \ref{Appendix:Longitudinal}). 

First of all, all the terms of the Kubo conductivity 
with the random current vertices are recovered. 
In addition, we find a new operator  
\be
-\frac{2\pi}{8\cdot 32  \Sigma^4} \Pi
\int dR\,
Tr\left[Q^\dagger(R)\vec{\nabla}Q(R)\sigma_3\right]
\cdot Tr\left[Q^\dagger(R)\vec{\nabla}Q(R)\sigma_3\right],
\label{Anomalo}
\ee
which has contributions from two different terms. 
The first one is 
obtained by expanding each Green's function in 
(\ref{Act:second}) at first order, $G_P = G_0 -iG_0PG_0$, and the second 
is derived by (\ref{SQ_P}). They are analogous to the components  
of the Kubo conductivity with regular and with 
random current vertices, respectively. 

\section{Effective non--linear $\sigma$-model}
\label{Effective NLSM}

In conclusion the final expression of the action of the 
transverse modes in the long-wavelength limit is
\bea 
S[Q] &=& \frac{2\pi}{32\Sigma^2}\sigma_{xx}
\int dR\, Tr\left(\vec{\nabla}Q(R) \cdot\vec{\nabla} Q(R)^\dagger\right)
\nonumber\\
&+& \int dR\, i\frac{E}{\omega_0}Tr\left(Q(R)\right)
- \frac{\omega}{2\omega_0}Tr\left(s_3 Q(R)\right)\nonumber\\
&-&\frac{2\pi}{8\cdot 32  \Sigma^4}\Pi
\int dR\,
Tr\left[Q^\dagger(R)\vec{\nabla}Q(R)\sigma_3\right]
\cdot Tr\left[Q^\dagger(R)\vec{\nabla}Q(R)\sigma_3\right],
\label{NLsM}
\eea
where we make use of the fact that, in the models we consider, 
$\sigma_{ij}= \delta_{ij}\sigma_{xx}$.
Since $Q(R)=Q_{sp}T(R)^2$, see Eq.(\ref{Q:matrix}), expressing 
$T(R)$ by means of $U(R)$ as in Eq.(\ref{T-unitary}),
the action at $E=\omega=0$ can also be written as 
\bea
S[U] &=& \frac{2\pi\sigma_{xx}}{16} \int dR\; 
Tr\left[ \vec{\nabla}U(R)^{-2} \cdot \vec{\nabla}U(R)^2\right]
\nonumber\\
&-& \frac{2\pi}{ 32}\frac{\Pi}{2}\int dR\; 
\left\{Tr\left[U(R)^{-2}\, \vec{\nabla}U(R)^2\right]\right\}^2,
\label{Effective:NLSM}
\eea
as anticipated in the section \ref{Summary}. 
As compared to the non-linear $\sigma$-model which is obtained 
in the absence of the particle-hole symmetry\cite{EL&K}, 
the above action differs 
first of all because of the symmetry properties of the matrix field $U(R)$, 
which describes now the Goldstone modes within the coset space 
${\rm U}(4m)/{\rm Sp}(2m)$. Moreover, it also differs for 
the last term of (\ref{NLsM}), which, in the general case, even if present, 
is not related to massless modes. An analogous term was 
originally obtained by Gade\cite{Gade} in a two sublattice model 
described by two on-site levels with a regular hopping of the 
form $H_{RR'}=t_{RR'}\sigma_1$, 
and a local time-reversal 
symmetry breaking random potential 
$H_{imp,R} = w_{1R}\sigma_1 + w_{2R}\sigma_2$, which we discuss 
in section \ref{Time-reversal symmetry breaking}. 

Although the action may be parametrized by a simple unitary field 
$U(R)$ as in (\ref{Effective:NLSM}), we prefer to work with the 
matrix $Q(R)$ which has the more transparent physical interpretation 
$Q(R)\sim \Psi(R)\overline{\Psi}(R)$.

Finally, it is important to notice that either $\sigma_{xx}$ and $\Pi$ 
have contributions from both the regular and the random current vertices. 
This implies that, even in the limit of strong disorder, in which the 
average hopping is negligible with respect to its fluctuations, these 
constants are finite and become of order unity\cite{O&W,Altland}.

\subsection{Gaussian Propagators}

At second order in $W$, the dispersion term 
\ben
&&\frac{2\pi\sigma_{xx}}{32\Sigma^2}\int dR\;
Tr\left({\vec \nabla} Q^\dagger{\vec \nabla} Q\right) 
\simeq 
-\frac{2\pi \sigma_{xx}}{32}\int dR\;
Tr\left({\vec \nabla} W {\vec \nabla} W\right) \\
&=&
\frac{2\pi \sigma_{xx}}{32}\int dR\;
4Tr\left({\vec \nabla} B {\vec \nabla} B^\dagger \right) 
+ 2 
Tr\left({\vec \nabla} A {\vec \nabla} A + 
{\vec \nabla} C {\vec \nabla} C\right),
\een
where $A$, $B$ and $C$ are defined through Eqs.(\ref{A:def}), 
(\ref{B:def}) and (\ref{C:def}). 
For the $B$'s we find the quadratic action
\[ 
\frac{\pi \sigma_{xx}}{2}\sum_{i=0}^4 
\sum_k \sum_{ab} k^2 B_{i,ab}(k)B_{i,ab}(-k),
\]
so that 
\be
\langle B_{i,ab}(k)B_{j,cd}(-k)\rangle = 
\delta_{ij}\delta_{ac}\delta_{bd}D(k),
\label{ActionB}
\ee
where
\be
D(k) = \frac{1}{\pi \sigma_{xx}} \frac{1}{k^2}.
\label{Dk}
\ee

For the $A$'s we have to take into account also the disconnected term:
\ben
&&-\frac{2\pi \Pi}{32\cdot 8 \Sigma^4}
\int dR\,
Tr\left[Q^\dagger(R)\vec{\nabla}Q(R)\sigma_3\right]
\cdot Tr\left[Q^\dagger(R)\vec{\nabla}Q(R)\sigma_3\right]\\
&&\simeq - \frac{2\pi \Pi}{64}
\int dR\,
Tr\left[\vec{\nabla}W_3\right]
\cdot Tr\left[\vec{\nabla}W_3\right]\\
&& = \frac{2\pi \Pi}{64}
\int dR\,
Tr\left[\vec{\nabla}A + \vec{\nabla}C \right]
\cdot 
Tr\left[\vec{\nabla}A + \vec{\nabla}C \right]\\
&& = \frac{2\pi \Pi}{16}
\int dR\,
Tr\left[\vec{\nabla}A_0 + \vec{\nabla}C_0 \right]
\cdot 
Tr\left[\vec{\nabla}A_0 + \vec{\nabla}C_0 \right].
\een 
The non vanishing propagators are 
\bea
\langle A_{0,ab}(k)A_{0,cd}(-k) \rangle &=&
D(k) 
\left(\delta_{ac}\delta_{bd}+\delta_{ad}\delta_{bc}\right)\nonumber\\
&-& D(k)\frac{\Pi}{\sigma_{xx}+\Pi m}\delta_{ab}\delta_{cd},
\label{A0}\\
\langle C_{0,ab}(k)C_{0,cd}(-k) \rangle &=&
D(k) 
\left(\delta_{ac}\delta_{bd}+\delta_{ad}\delta_{bc}\right)\nonumber\\
&-& D(k)\frac{\Pi}{\sigma_{xx}+\Pi m}\delta_{ab}\delta_{cd},
\label{C0}\\
\langle A_{0,ab}(k)C_{0,cd}(-k) \rangle &=&
- D(k)\frac{\Pi}{\sigma_{xx}+\Pi m}\delta_{ab}\delta_{cd},
\label{AC0}
\eea
where $m$ is the number of replicas, while for $i=1,2,3$
\bea
\langle A_{i,ab}(k)A_{i,cd}(-k) \rangle &=&
D(k)
\left(\delta_{ac}\delta_{bd}-\delta_{ad}\delta_{bc}\right)
\label{Ai}\\
\langle C_{i,ab}(k)C_{i,cd}(-k) \rangle &=&
D(k)
\left(\delta_{ac}\delta_{bd}-\delta_{ad}\delta_{bc}\right)
\label{Ci}.
\eea

Notice that the particular symmetry of the two-sublattice model 
leads to additional diffusive modes in the retarded-retarded and 
advanced-advanced channels, which are not massless in the standard 
case\cite{EL&K}.

\subsection{Physical Meaning of $\Pi$}
\label{Physical Meaning of Pi}
Let us introduce an external source which couples to the 
staggered density of states, which is accomplished by 
adding to the action a term
\[
\int dR\, \overline{\Psi}_R s_3\sigma_3 \hat{\lambda}(R) \Psi_R,
\]
where, in the replica space, the source 
$\hat{\lambda}_{\alpha,\beta}= \lambda_\alpha \delta_{\alpha,\beta}$. The 
fluctuations of the staggered density of states is obtained by the 
derivative of the partition function with respect, for instance, to 
$\lambda_\alpha$ and $\lambda_\beta$, with $\alpha\not =\beta$.    
Inserting the source term in the action, and integrating over the 
Grassmann fields after introducing the matrix $Q$, leads to the following
expression of the staggered density of states fluctuation in terms of 
$Q$:
\be
F(R,R') = \frac{1}{\pi^2\omega_0^2}
\langle Tr\left[Q_{\alpha\alpha}(R)s_3\sigma_3\right] 
Tr\left[Q_{\beta\beta}(R')s_3\sigma_3\right] \rangle.
\ee
The gaussian estimate of the above correlation function at momentum $k$ 
is given by 
\be
F(k) = -16\frac{\Sigma^2}{\pi^2 \omega_0^2} \langle 
\left[A_{0,\alpha\alpha}(k) + C_{0,\alpha\alpha}(k)\right] 
\left[A_{0,\beta\beta}(-k) + C_{0,\beta\beta}(-k)\right] \rangle 
= \frac{64\Sigma^2}{\pi^2 \omega_0^2} D(k)\frac{\Pi}{\sigma_{xx}+\Pi m}.
\ee
Therefore, $\Pi$ is directly related to the singular behavior of the 
staggered density of states fluctuations.

\section{Renormalization Group}
\label{Renormalization Group}

In this section, we will apply the Wilson--Polyakov Renormalization Group (RG)
procedure\cite{RGref,EL&K} to analyse the scaling behavior of the 
action. 
Indeed, some of the calculations which we present are redumdant, given the 
proof by Gade and Wegner that the $\beta$-function is zero\cite{W&G} 
(see Appendix \ref{App:Symmetries}). 
Nevertheless, other results besides the conductance $\beta$-function 
are important, so that we describe the whole RG procedure.

\subsection{RG equations}

In the spirit of Wilson--Polyakov RG approach\cite{EL&K,RGref}, we assume that 
\[
T(R) = T_f(R)T_s(R),
\]
where $T_f$ involves fast modes with momentum $q\in [\Lambda/s,\Lambda]$, 
while $T_s$ involves slow modes with $q\in [0,\Lambda/s]$,   
being $\Lambda$ the higher momentum cut-off, and the rescaling factor $s>1$.
Within an $\epsilon$--expansion, where $\epsilon=d-2$, we define
\[
\int_{\Lambda/s}^\Lambda \frac{d\vec{k}}{(2\pi)^d} D(k)
\equiv L = \frac{1}{4\pi^2\sigma_{xx}}\ln s + {\rm O}(\epsilon).
\]

It is straightforward to show the following result
\bea
&& Tr\left[\vec{\nabla}Q^\dagger \vec{\nabla} Q\right] =
Tr\left[\vec{\nabla}Q_f^\dagger \cdot \vec{\nabla} Q_f\right]
\nonumber\\
&& +2Tr\left[\vec{D}_s\sigma_1 Q_f \vec{D}_s Q_f^\dagger \sigma_1\right]
-2\Sigma^2 Tr\left[\vec{D}_s \vec{D}_s \right]\nonumber\\
&&+4Tr\left[\vec{D}_s Q_f^\dagger \vec{\nabla} Q_f \right],
\label{RG:dQdQ}
\eea
where $Q_f = \tilde{T}_f^\dagger Q_{sp} T_f$ and 
$\vec{D}_s = T_s\vec{\nabla}T_s^\dagger$. Moreover,
\bea
&&\frac{1}{\Sigma^4}Tr\left[Q^\dagger \vec{\nabla} Q\sigma_3\right] 
\cdot Tr\left[Q^\dagger \vec{\nabla} Q\sigma_3\right] \nonumber\\
&&= 
Tr\left[\left(\vec{\nabla}W_s + \vec{\nabla} W_f\right)\sigma_3\right]
\cdot Tr\left[\left(\vec{\nabla}W_s + \vec{\nabla} W_f\right)\sigma_3\right].
\label{RG:QdQ}
\eea
Since the fast and slow modes live in disconnected regions of momentum 
space, only the stiffness term (\ref{RG:dQdQ}) generates corrections. 
By expanding the terms 
coupling slow and fast modes up to second order in $W_f$, the stiffness 
generates an action term for the slow modes which, after averaging over 
the fast ones, is 
\be
\frac{2\pi\sigma_{xx}}{32\Sigma^2} \int dR\, 
Tr\left[ \vec{\nabla}Q_s^\dagger \vec{\nabla}Q_s \right]
+ \langle F_1\rangle_f - \frac{1}{2} \langle F_2^2 \rangle_f,
\ee
where 
\bea
F_1 &=& \frac{2\pi\sigma_{xx}}{32\Sigma^2} \int dR\, 
-2 Tr\left[ \vec{D}\sigma_1 Q_{sp} W_f \vec{D} W_f Q_{sp}
\sigma_1\right]
\nonumber\\ 
&&+2 Tr\left[ \vec{D}\sigma_1 Q_{sp} \vec{D} Q_{sp} 
\tilde{W}_f^2
\sigma_1\right],
\label{F1}
\eea
and 
\be
F_2 = 4 \frac{2\pi\sigma_{xx}}{32}
\int dR\, Tr\left[\vec{D}W_f\vec{\nabla}W_f\right].
\ee
The explicit evaluation of these terms is outlined in Appendix 
\ref{Appendix:RG}. Here we just give the final results. The 
Kubo conductivity is renormalized according to 
\be
\sigma_{xx} \to \left(1 - 4Lm \right)\sigma_{xx},
\label{RG:sigma_xx}
\ee
while the $\Pi$ factor 
\be
\Pi \to \Pi + 4L\sigma_{xx}.
\ee

For what it regards the renormalization of $E$ and $\omega$, we notice that 
\ben
Q  &=& Q_{sp} T_s T_f^2 T_s\\
&\simeq& Q_{sp} T_s^2 + Q_{sp} T_s W_f T_s + 
\frac{1}{2}Q_{sp} T_s W_f^2 T_s.
\een
Since the slow and fast degrees of freedom are defined in 
different regions of momentum space, only the second term is 
relevant. By means of Eq.(\ref{App:WW}), we
find that 
\be
Q \to Q_s \left\{ 
1 + \frac{1}{2}\left(2 - 8m 
+ \frac{\Pi}{\sigma_{xx}+m\Pi}\right)L 
\right\} 
\label{RG:Q}
\ee
This leads to similar corrections to $E$ and $\omega$, which will 
have the same scaling behavior. 

Finally, to describe the cross-over behavior in the presence of 
symmetry breaking terms, we also need the scaling behavior of 
the operator $Tr\left[Q(R)^2\right]$. We find  that 
\bea
\langle Tr Q^2 \rangle_f &=& 
\langle Tr\left[ Q_{sp} T_s T_f^2 T_s Q_{sp} T_s T_f^2 T_s\right] \rangle_f 
\nonumber\\
&=&
\left[1+2\left(2 -4m + \frac{\Pi}{\sigma_{xx}+m\Pi}\right)L\right]
Tr Q_s^2 - L \left( Tr Q_s \right)^2.
\label{RG:Q^2}
\eea

To implement the RG, we have to rescale the momenta in order to 
recover the original range $[0,\Lambda]$. This is accomplished by 
the transformation $k\to k/s$, or, equivalently, $R\to Rs$. Therefore, 
the stiffness as well as the fluctuation terms acquire a scaling factor 
$s^\epsilon \simeq 1 + \epsilon\ln s$, while the $E$ and $\omega$ 
terms a factor $s^d$. Hence, after defining 
$t=1/(4\pi^2\sigma_{xx})$, $c=1/(4\pi^2\Pi)$, and $\lambda$ the coupling 
constant of the operator $Tr Q^2$,  
we get the following $\beta$-functions 
\bea
\beta_t &=& -\epsilon t + 4mt^2,\label{beta:t}\\
\beta_c &=& -\epsilon c - 4 c^2,\label{beta:c}\\
\beta_E &=& dE + E\frac{t}{2} \left(2 + \frac{t}{c+mt}\right),
\label{beta:E}\\
\beta_\lambda &=& d\lambda + 2 \lambda t \left(2 + \frac{t}{c+mt}\right).
\label{beta:lambda}\\
\eea 

At finite energy, $E\not =0$, we may use a two-cutoff scaling 
approach\cite{Gade}. 
Namely, we can follow the previous RG equations up to a 
cross-over scale, $s_{cross}=s(E)$, at which the 
energy has flowed to a value $E_0$ of order $\Sigma$, which 
plays the role of the high-energy cut-off in the theory.  
Above this scale, we must neglect all contributions  
coming from the $W_3$ modes, which acquire a mass. That is, we 
must abandon the RG equations (\ref{beta:t})--(\ref{beta:lambda}), and let 
the coupling constants flow in accordance with the standard RG equations, 
which amount only to a renormalization of $t$ according to    
\be
\beta_t = -\epsilon t + t^2.
\label{beta:tE}
\ee
If, by integrating (\ref{beta:tE}), the inverse conductance $t(s)$ flows to 
infinity, signalling an insulating behavior, then we can define 
a localization length $\xi_{loc}(E)$ as the scale 
at which $t$ has grown to a value of order unity. 
 
\subsection{RG in $d=2$}

In $d=2$, the solution of the RG equations for $t$, $c$, and $E$ is 
\ben
t(s)&=&t(1)=t_0,\\
\frac{1}{c(s)} &=& \frac{1}{c(1)} + 4\ln s=
\frac{1}{c_0} + 4\ln s,\\
\ln\frac{E(s)}{E} &=& 
\left[2 + \frac{t_0}{2} \left(2+\frac{t_0}{c_0}\right)\right]\ln s
+\frac{1}{2} t_0^2 \ln^2 s
\een
At finite energy, the crossover length, 
$s(E)$, to the standard, non particle-hole 
symmetric, model, which, as previously discussed, is defined through 
$E[s(E)]=E_0\sim \Sigma$, is given by     
\be
s(E) = 
\exp\left\{\frac{1}{2A}\left(\sqrt{B^2+4A\ln\frac{E_0}{E}} - B\right)
\right\},
\label{xiloc}
\ee
being 
\[
A = \frac{1}{2} t_0^2,\;\;
B = 2 + \frac{t_0}{2}\left(2+\frac{t_0}{c_0}\right).
\] 
Above $s(E)$, $t$ flows according to Eq.(\ref{beta:tE}) 
with $\epsilon=0$, hence it grows to infinity,  
implying that the wavefunctions are localized for any $E\not =0$. 
The localization length $\xi_{loc}(E)\sim s(E)$, apart from a 
multiplicative factor which is $\sim \exp[(1-t_0)/t_0]$ if 
$t_0\ll 1$. We see that, for 
\[
E\gg E_0 \exp\left\{
-\frac{1}{2t_0^2}
\left[2 + \frac{t_0}{2} \left(2+\frac{t_0}{c_0}\right)\right]^2\right\},
\]
the localization length has a power law behavior, namely
\[
\xi_{loc}(E) \propto \left(\frac{E_0}{E}\right)^{\frac{1}{B}},
\]
otherwise, at very small energy, it diverges slower, 
\[
\xi_{loc}(E) \propto \exp\sqrt{\frac{\ln(E_0/E)}{A}}.
\]

The density of states renormalizes like 
\[
\ln\frac{\rho(s)}{\rho_0} = 
\frac{t_0}{2}\left(2+\frac{t_0}{c_0}\right)\ln s
+\frac{1}{2}t_0^2\ln^2 s.
\]
At finite energy, the density of states flows until $s<S(E)$, after which 
it stays constant. This 
implies that the renormalized value is obtained by 
\[
\ln\frac{\rho(E)}{\rho_0} = 
\ln\frac{\rho(s(E))}{\rho_0} = \ln\frac{E_0}{E} -2\ln s(E),
\]
leading to  
\be
\rho(E)=\rho_0 \left(\frac{E_0}{E}\right)\frac{1}{s(E)^2}
\simeq \rho_0 \left(\frac{E_0}{E}\right) 
\exp\left(-\sqrt{\frac{4\ln(E_0/E)}{A}}\right),
\ee
the last equality valid at small energy. 

\subsection{RG in $d=3$}
In $d=3$, 
\ben
t(s)&=& t_0s^{-1},\\
c(s) &=& \frac{c_0 s^{-1}}
{1 + 4c_0 -4c_0 s^{-1}},\\
\ln \frac{E(s)}{E} &=& 
3\ln s + \frac{t_0}{2}\left(2+\frac{t_0}{c_0}+4t_0\right)
\left(1-\frac{1}{s}\right) - t_0^2\left(1-\frac{1}{s^2}\right),
\een
hence both $t$ and $c$ running variables vanish for $s\to\infty$. The 
cross-over length diverges, as we approach $E=0$, approximately 
like 
\be
s(E)\simeq  \left(\frac{E_0}{E}\right)^{\frac{1}{3}}
{\rm e}^{-\frac{t_0}{6}\left(2+\frac{t_0}{c_0}+t_0\right)}.
\label{xiloc3d}
\ee
Hence, the density of states 
\be
\rho(E) = \rho_0 \left(\frac{E_0}{E}\right)\frac{1}{s(E)^3}
\simeq \rho_0 {\rm e}^{\frac{t_0}{2}\left(2+\frac{t_0}{c_0}+t_0\right)},
\ee
saturates at $E=0$ to a value exponentially increased in $t_0$ with 
respect to the bare $\rho_0$. 

At $E\not = 0$, the inverse conductance above $s_{cross}=s(E)$ flows 
according to (\ref{beta:tE}) with $\epsilon=1$ and 
boundary condition $t(s_{cross})=t_0/s_{cross}$.  
We find that, if
\[
t_0 < s_{cross}, 
\]
the system is metallic, otherwise it is 
insulating, with a localization length 
\[
\xi_{loc}(E) \sim \frac{s(E)}{t_0 - s(E)}.
\]
This results implies that, for any amount 
of disorder, sufficiently close to $E=0$, all eigenfunctions 
are delocalized, in agreement with recent numerical results\cite{Cain}.
However, if the disorder is gaussian, 
as we assumed, the random hopping model with zero regular hopping 
seems to be characterized by an inverse Drude conductance, $t_0$,   
which is an increasing function of $|E|$, being smaller 
than the critical value $\epsilon$ at $E=0$ (see also Ref.\cite{O&W}). 
In this case, the presence of a finite 
mobility edge in $d=3$, even for zero regular hopping, would not depend 
crucially upon the intermediate RG flow in the vicinity to the band center.   
Nevertheless, we expect that $t_0$ at $E=0$ varies for different kinds of 
disorder, and eventually it may become greater than unity. 
In this case, it is just the vicinity to the band center which 
makes it possible a finite mobility edge.

\section{On site disorder}
\label{On site disorder}

In this and in the following section, we analyse various symmetry breaking 
terms, which, in the two sublattice representation, contain $\sigma_0$ 
and $\sigma_3$, hence spoiling Eq.(\ref{Summary:condition}).  

We start by adding an onsite disorder 
\[
\delta S_{imp} = \sum_R u_{1,R}\, \cc_{1,R}\ca_{1,R}
+u_{2,R}\, \cc_{2,R}\ca_{2,R} = 
\sum_R \cc_R \left(\frac{u_{1,R}+u_{2,R}}{2}\sigma_0 
+ \frac{u_{1,R}-u_{2,R}}{2}\sigma_3\right) \ca_R,
\]
where $\langle u_{i,R}\rangle = 0$ and 
$\langle u_{i,R}u_{j,R'}\rangle =\delta_{ij}\delta_{RR'}v^2$. 
Within the path integral, this term becomes, once average over disorder 
is performed,
\ben
\delta S_{imp} &=& \frac{v^2}{2V}\sum_q 
Y_{1,q}^{\alpha\beta}Y_{1,-q}^{\beta\alpha}
+ Y_{2,q}^{\alpha\beta}Y_{2,-q}^{\beta\alpha}\\
&=& \frac{v^2}{4V}\sum_q 
Y_{0,q}^{\alpha\beta}Y_{0,-q}^{\beta\alpha}
+ Y_{3,q}^{\alpha\beta}Y_{3,-q}^{\beta\alpha}.
\een
By adding this term to (\ref{Path:Simp2}), we get 
\bea
S_{imp}+\delta S_{imp} &=& 
\frac{1}{4V}\sum_q (\omega_q+v^2)Tr\left(
Y_{0,q}Y_{0,-q}\right) \nonumber\\
&& -(\omega_q-v^2)Tr\left(Y_{3,q}Y_{3,-q}\right).
\label{Onsite}
\eea
If we assume that the onsite disorder is weak, i.e. $\omega_q>v^2$ 
at small $q$, the consequence is that the $Q$ free action becomes
\bea
S^0_{imp} &=& \frac{1}{V}
\sum_q \frac{1}{\omega_q+v^2}Tr\left[Q_{0,q}Q_{0,-q}\right]
+\frac{1}{\omega_q-v^2}Tr\left[Q_{3,q}Q_{3,-q}\right]\nonumber\\
&=& \frac{1}{V}
\sum_q \frac{1}{2(\omega_q+v^2)}Tr\left[Q_{q}Q_{q}^\dagger\right]
+\frac{2v^2}{\omega_q^2-v^4}Tr\left[Q_{3,q}Q_{3,-q}\right].
\label{OnsiteS0}
\eea
Therefore, the on site disorder introduces a mass in the $Q_3$ propagators. 
Specifically, since $
2i Q_3\sigma_3 = Q - Q^\dagger$, 
the mass term can be written as 
\[
-\frac{v^2}{4(\omega_q^2-v^4)}Tr\left[
\left(Q_q - Q^\dagger_q\right)
\left(Q_q - Q^\dagger_q\right)
\right].
\]
Close to the saddle point, $QQ^\dagger=Q_{sp}^2$, and, for small $q$, 
we get 
\[
-\frac{v^2}{4(\omega_0^2-v^4)}\int dR\; Tr\left[
Q(R)Q(R) + Q(R)^\dagger Q(R)^\dagger
\right],
\]
which, at second order in $W$, reads   
\bea
&&-\frac{v^2\Sigma^2}{2(\omega_0^2-v^4)}\int dR\; Tr\left[
W(R)W(R) + W(R)s_3W(R)s_3\right]\nonumber \\
&&
= -\frac{2v^2\Sigma^2}{\omega_0^2-v^4}\int dR\; Tr\left[
W_3(R)W_3(R) \right].
\label{Onsitemass}
\eea
In the presence of this term, we could proceed, as before,  
in the framework of 
two cutoff scaling theory. That is, we apply the previous RG equations 
until the above term becomes of the order $\Sigma\rho_0$, i.e. up to the scale 
which, by Eq.(\ref{beta:lambda}), is
\be
s_{cross} = 
\exp\left\{\frac{1}{2A}\left(\sqrt{B^2+4A\ln\lambda} - B\right)
\right\}, 
\label{scross}
\ee
in $d=2$, where $A=2t_0^2$ and $B=2 + t_0(2+t_0/c_0)$, 
while $s_{cross}\simeq \lambda^{1/d}$ in $d>2$, where 
\[
\lambda \propto \frac{\omega_0}{v^2},
\]
in the limit of small $v$.  
Above this scale, the $W_3$ propagator gets fully massive, 
and the inverse conductivity flows with the RG equation (\ref{beta:tE}).  
In $d=2$, this implies that, ultimately, the system gets localized, 
although the density of states has increased in the first stage of the RG. 

\section{Same--Sublattice Regular Hopping}
\label{Same--Sublattice Regular Hopping}

We can also introduce a particle--hole symmetry breaking term, by adding to 
the Hamiltonian a regular term connecting same sublattices, e.g.
\[
\delta H_{RR'} = t^{(0)}_{RR'}\sigma_0 + 
t^{(3)}_{RR'}\sigma_3 \to \delta H_k = t^{(0)}_k\sigma_0 +
t^{(3)}_k \sigma_3.
\]
Expanding the action in $\delta H$, after integrating over the 
Nambu spinors, we get an additional term
\be
\delta S[Q] = \frac{1}{2}Tr\left[\tilde{T}\,\delta H\, T^\dagger
\, G\right].
\label{106}
\ee  
We define
\ben
\frac{4}{\omega_0} 
\lambda_0 Q_{sp} &\equiv& i \frac{1}{V}\sum_k t^{(0)}_k G(k), \\
\frac{4}{\omega_0} 
\lambda_3 Q_{sp} &\equiv& -i \frac{1}{V}\sum_k t^{(3)}_k 
G(k),
\een 
so that (\ref{106}) becomes
\be
-i \frac{4}{\omega_0} Tr\left[Q\left(\lambda_0\sigma_0
+\lambda_3\sigma_3\right)\right].
\label{PHB}
\ee
The $\lambda_0$-term acts like an energy term. This implies that, if 
we just shift the chemical potential, we do recover the same 
scenario as in the absence of this term and at $E=0$. On the contrary, the  
$\lambda_3$-term is always a relevant perturbation, whose strength increases 
under RG iteration as the energy $E$. We can define  
a crossover scale $s_{cross}$, which has the same
expression as $s(E)$ in (\ref{xiloc}) and (\ref{xiloc3d}), for 
$d=2$ and $d=3$, respectively, provided $E\to \lambda_3$. 
Above this scale, the $W_3$ modes get fully massive and their contribution 
to the RG flow drops out. 

Sometimes $t^{(3)}_{RR'}=0$, as, for instance, for  
next-nearest neighbor hopping in a square lattice. In this case, 
$\lambda_3=0$ and   
we need to evaluate the second order term 
\[
\delta S[Q] = \frac{1}{4}Tr\left[
\tilde{T}\,\delta H\, T^\dagger
\, G \, \tilde{T}\,\delta H\, T^\dagger
\, G\right].
\]
If we define $F(R)=
\tilde{T}(R)T(R)^\dagger$, which contains either $\sigma_0$ and $\sigma_3$,
this term can be 
written, at long wavelengths, as 
\[
\delta S[Q] = \frac{1}{4}\sum_k 
\left(t^{(0)}_k\right)^2 
Tr\left[ F_{q} G(k) F_{-q} G(k)\right].
\]
By introducing,  
\[
\Sigma^{pq}_i = \frac{1}{V} \sum_k \left(t^{(0)}_k\right)^2
\, Tr\left[\sigma_i G^p \sigma_i G^q\right],
\]
where $p,q=\pm$, the following results hold
\[
\Sigma^{pq}_0 = \frac{1}{2}(1-pq)\frac{4C}{\omega_0},\;\;
\Sigma^{pq}_3 = -\frac{1}{2}(1+pq)\frac{4C}{\omega_0}, 
\] 
where $C$ is a constant of dimension 
energy square, with order of magnitude given by the typical value of 
$\left(t^{(0)}_k\right)^2$ close to the surface corresponding to $E=0$. 
Therefore we can write, 
\bea
\delta S[Q] &=& \frac{C}{2\omega_0} 
\int dR\, Tr\left[ F(R)\tilde{F}(R) 
- F(R)s_3\sigma_1\tilde{F}(R)s_3\sigma_1\right]\nonumber \\
&=& {\rm const.} + \frac{C}{2\omega_0\Sigma^2} 
\int dR\, Tr\left[ Q(R)^2 \right], 
\label{uno}
\eea
which is a mass term for 
the $W_3$ propagators, similar to (\ref{Onsitemass}). Therefore, 
a same-sublattice hopping introduces a cross-over length analogous 
to (\ref{scross}), with 
\[
\lambda \propto 
\frac{\Sigma^2}{C}.
\]
Above this scale, the contribution of the $W_3$ modes 
to the RG flow has to be dropped out.   

\section{Time-reversal symmetry breaking}
\label{Time-reversal symmetry breaking}

If the random hopping breaks time-reversal symmetry, i.e.
\[
H_{imp} = \sum_{RR'} \tau^{12}_{RR'}\cc_R\ca_{R'}
+ H.c.,
\]
with both 
real and imaginary part of $\tau^{12}_{RR'}$ gaussian distributed,   
after averaging, the impurity action can be written as  
\be
S_{imp} = \frac{1}{V}\sum_q W_{-q} 
Tr\left[ X_{1,0,q}X_{2,0,-q}+X_{1,3,q}X_{2,3,-q}\right],
\ee
where 
\[
X^{\alpha\beta}_{1,0,R}= \overline{\Psi}^\alpha_{1R}\tau_0
\Psi^\beta_{1R},\;\;
X^{\alpha\beta}_{1,3,R}= \overline{\Psi}^\alpha_{1R}\tau_3
\Psi^\beta_{1R},
\]
with the indices $\alpha$ and $\beta$ running only over 
the replicas and the advanced/retarded components. 
This implies that the manifold in which $Q$ varies contains in 
this case only $\tau_0$ and $\tau_3$ components. Indeed, as in the 
time reversal invariant case we are able to parametrize the 
$8m\times 8m$ matrix $T$ in terms of a 
$4m\times 4m$ matrix $U\in {\rm U}(4m)/{\rm Sp}(2m)$ 
[see Eq.(\ref{T-unitary})], similarly, 
without time-reversal symmetry, $T$ can be parametrized by 
means of a $2m\times 2m$ matrix $U\in {\rm U}(2m)$, in agreement with 
Gade\cite{Gade}. 
The effective non-linear $\sigma$-model is not modified, but 
the expressions (\ref{BBdag}), 
(\ref{BB}), (\ref{AA}),  
(\ref{CC}), and (\ref{AC}) have to be substituted by 
\bea
\langle B_{ab}P_{bc}B_{cd}^\dagger\rangle &=& 
 2D(k)\delta_{ad} Tr\left(P_0\right),\label{BBdag-noT}\\
\langle B_{ab}P_{bc}B_{cd}\rangle &=& 0,\label{BB-noT}\\
\langle A_{ab}P_{bc}A_{cd}\rangle &=& 
2D(K)\delta_{ad}Tr\left(P_0\right)
- D(k)\frac{\Pi}{\sigma_{xx}+\Pi m} P_{ad},\label{AA-noT}\\
\langle C_{ab}P_{bc}C_{cd}\rangle &=& 
2D(K)\delta_{ad}Tr\left(P_0\right)
- D(k)\frac{\Pi}{\sigma_{xx}+\Pi m} P_{ad},\label{CC-noT}\\
\langle A_{ab}P_{bc}C_{cd}\rangle &=& 
- D(k)\frac{\Pi}{\sigma_{xx}+\Pi m} P_{ad}.
\eea
where $P=P_0+iP_3\tau_3$. Hence, the RG equations at $m=0$ 
are, in this case, 
\bea
\beta_t &=& -\epsilon t,\label{beta:t-noT}\\
\beta_c &=& -\epsilon c - 2 c^2,\label{beta:c-noT}\\
\beta_E &=& dE + \frac{Et^2}{2c},
\label{beta:E-noT}
\eea 
which coincide with those obtained by Gade\cite{Gade}.

\section{Discussion and comparison with the standard localization theory}

In this section, we summarize the main differences between the 
model (\ref{Summary:NLSM}) and the standard non-linear 
$\sigma$-model which is derived in the theory of Anderson localization
\cite{Wegnersigma,EL&K}, placing particular emphasis on the 
properties of the $Q$-matrix. The specific form of the off-diagonal disorder 
we consider,  
which only couples one sublattice to the other, leads, via the 
Hubbard-Stratonovich decoupling, to the introduction of a space-varying 
$8m\times 8m$ complex $Q$-matrix, $Q=Q_0\sigma_0 + i Q_3 \sigma_3$. Here,  
$Q_0$ and $Q_3$ are $4m\times 4m$ hermitean matrices, of which  
matricial structure refers to the retarded/advanced, 
spinor particle/hole and $m$ replica components. 
Contrary to the standard case, $Q$ is not hermitean. 

The evaluation of the saddle point, 
$Q_{sp} = \sigma_0\tau_0 s_3$ (section\ref{Saddle Point}), 
as well as the derivation of the efective action  
(section \ref{Effective Action}) 
are analogous to the standard case \cite{Wegnersigma,EL&K}. (We recall that 
$\sigma_i$, $s_i$ and $\tau_i$ indicate the Pauli matrices, 
including the unit matrix, acting on sublattice, advanced/retarded 
and spinor components, respectively.) 
The non-linear $\sigma$-model, Eq.(\ref{NLsM}), is obtained by 
integrating out the longitudinal massive $Q$-fluctuations and only  
keeping the transverse soft modes. The real novelty with respect to 
localization theory is not in the structure of the effective action. 
Indeed, the new term in (\ref{NLsM}), namely 
\[
\Pi \int dR 
\left[Tr \left(Q(R)^\dagger \vec{\nabla} Q(R) \sigma_3\right)\right]^2,
\] 
even if present, would be irrelevant in the standard case.    
On the contrary, the essential difference, as expected, 
lies in the ensamble spanned by the 
soft modes at the particle-hole symmetry point $E=0$.  
We get $Q_{Soft} = \tilde{T}^{-1} Q_{sp} T$, where the unitary 
matrix $T$ only contains $\sigma_0$ and $\sigma_3$, 
\[
T=\exp\left[\frac{W_0\sigma_0+W_3\sigma_3}{2}\right],
\]
and 
\[
\tilde{T}= \sigma_1 T \sigma_1 = \sigma_2 T \sigma_2 = 
\exp\left[\frac{W_0\sigma_0-W_3\sigma_3}{2}\right].
\] 
These expressions derive by the conditions (\ref{Summary:condition}), 
which fully specify the model, as shown in section \ref{Symmetries}.
In that section, we also showed that the ensamble can be expressed in 
terms of unitary $4m\times 4m$ matrices 
\[
U = \exp\left[\frac{W_0+W_3}{2}\right],
\]
as argued by Gade and Wegner\cite{W&G}. Selecting the subset which leaves 
the saddle point invariant gives Eq.(\ref{Summary:NLSM}) with 
$U\in {\rm U}(4m)/{\rm Sp}(2m)$. In terms of $T$, the condition 
$\tilde{T}^{-1} Q_{sp} T \not = Q_{sp}$, leads to the requirements  
$[W_0,s_3]\not = 0$ and $\{W_3,s_3\}\not = 0$, which implies that 
$W_0$ is off-diagonal in the energy retarded/advanced space (as in the 
standard localization theory), while $W_3$ is diagonal. 
In other words, the omogeneous and staggered modes, $W_0$ and $W_3$, 
respectively, have different structure in the energy space. 
The energy diagonal $W_3$-modes betray the presence, at $E=0$, of 
diffusive poles in the disorder averaged products of retarded 
and advanced Green's functions, 
$\overline{G_R G_R}$ and $\overline{G_A G_A}$, with 
$G_{R,A}= \left(-H \pm i0^+\right)^{-1}$. In the localization theory 
\cite{Wegnersigma,EL&K}, only the mixed products $\overline{G_R G_A}$ have a 
singular behavior. This explains why singular corrections to the 
density of states (which involves connected diagrams with same energy Green's 
functions) 
are present in the two-sublattice model, while they are absent in the 
localization theory. 

In a square lattice, the energy diagonal modes have the transparent 
meaning of density fluctations with wave-vector $q$ nearby the 
nesting vector $G=(\pi,\pi,\dots)$, see Appendix A. Indeed
\ben
Q_3(q) &\sim& \sum_{R\in A} {\rm e}^{-iqR} 
\left(\Psi_{1R}\overline{\Psi}_{1R} 
- \Psi_{2R}\overline{\Psi}_{2R} \right) \\
&=& \sum_{R\in A,B} {\rm e}^{-i(q+G)R} \Psi_{R}\overline{\Psi}_{R}
= Q(q+G),
\een
where $A$ and $B$ label the two sublattice, and, for $R\in A$, 
we have taken by definition $\Psi_{1R}= \Psi_R$ and 
$\Psi_{2R}=\Psi_{R+a\hat{x}}$, being $a$ the lattice spacing and 
$\hat{x}$ the unit vector in the $x$-direction. As soon as $E\not =0$, 
nesting is not more important and indeed $Q_3$ becomes massive. 

Finally, because of $\overline{G_R G_R}$ and $\overline{G_A G_A}$, 
also the conductance acquires other corrections with respect 
to standard localization theory. Indeed, these corrections 
add to give a vanishing $\beta$-function for $\sigma_{xx}$, as first 
indicated by Gade and Wegner \cite{W&G}. 
We have explained in the Introduction section (see also Appendix E) that 
this is a consequence of a simple abelian gauge symmetry generated 
by Eq.(\ref{Summary:condition}) at the particle-hole symmetry point $E=0$. 
Similarly to the results for the density of states, this behavior 
of the $\beta$-function is at odds with the standard theory.

\section{Conclusions}
\label{Conclusions}

In this work, we have derived the effective non-linear $\sigma$-model of 
a disordered electronic system on a generic bipartite lattice. This model, 
if the hopping matrix elements as well as the disorder only couple 
one sublattice with the other, shows an interesting behavior close to the band 
center, i.e. to the particle-hole symmetry point. Namely, the 
wave-functions are always delocalized at the band center, in any 
dimension.  
By a Renormalization Group (RG) analysis, in the framework of 
an $\epsilon$-expansion, $\epsilon=d-2$, we have found that the 
quantum corrections to the conductivity vanish if the chemical 
potential is exactly at the band center, thus implying a metallic 
behavior. In two dimensions, in particular, the 
Kubo conductivity flows to a fixed value by iterating the RG. 
On the contrary, we have found that the 
staggered density of states fluctuations, which 
are controlled by a new parameter in the 
non-linear $\sigma$-model, are singular.  This result is reminiscent 
of what it is found in equivalent 
one-dimensional models. In fact, models of disordered spinless fermions  
in one-dimension can be mapped, by a Jordan-Wigner transformation, 
onto disordered spin chains. In many cases, it is known that, in spite 
of the presence of disorder, these spin chain models display critical 
behavior, as shown in great detail by D. Fisher for 
random Heisenberg antiferromagnets and random transverse-field 
Ising chains\cite{Fisher}. 
Indeed, as pointed out by Fisher, the staggered spin fluctuations, in a 
random antiferromagnetic chain,  
also display critical behavior in the form of a power law decay, 
$(-1)^R\overline{\langle S(R)\rangle\langle S(0)\rangle}\sim R^{-2}$,
where the bar indicates impurity 
average.
Since the staggered spin-density corresponds to the staggered density of the 
spinless fermions, this result is consistent with the outcome of our 
analysis, which further suggests that a similar scenario generally holds  
in such models. 
Moreover, as in 
one-dimension\cite{T&C,E&R,Fisher}, we find that the density of states 
is strongly modified by the disorder at the band center, and it actually 
diverges in $d=2$. 
In reality, a random Heisenberg chain, away from the $XXZ$ limit, maps 
onto a spinless fermion random hopping model in the presence of a random 
nearest-neighbor interaction. However, even in the presence of 
this additional interaction, the Hamiltonian has still the 
abelian gauge-like symmetry 
described in Appendix \ref{App:Symmetries}, which is at the origin of the 
delocalization of the band center state. This observation is also 
compatible with Fisher's result that the physical behavior does not 
qualitatively change upon moving away from the XXZ limit towards the 
isotropic XXX Heisenberg point.    

Many of the results which we have derived were already known. 
The existence of delocalized states at the band center of a two-sublattice 
model was argued already in 1979 by Wegner\cite{Wegner,O&W}. The 
effective non-linear $\sigma$-model when the disorder breaks 
time-reversal invariance, as well as the RG equations, have earlier been 
derived by Gade\cite{Gade}, although in a particular two-sublattice model. 
The extension to disordered systems with time-reversal symmetry was 
later on argued by Gade and Wegner\cite{W&G}. Finally, 
random flux models and disordered Dirac fermion models have recently been 
the subject of an intensive theoretical study\cite{Fukui,Altland,Ludwig}, 
for their implications to a variety of different physical problems. 

In spite of that, our analysis has several novelties with 
respect to earlier studies. First of all, the two-sublattice model 
which we study is quite general. Secondly, the physical interpretation of the 
parameters which appear in the non-linear $\sigma$-model is quite  
transparent. Thirdly, the explicit derivation of the RG equations 
with time-reversal invariance is presented.       

\section{Acknowledgments}
We are grateful to A. Nersesyan, V. Kratsov, E. Tosatti, and Yu Lu 
for helpful discussions and comments. 

\appendix

\section{Specific examples}
\label{App:Specific examples}

As an example, we consider a tight binding model with 
only nearest neighbor hopping on a square and honeycomb lattice. 
 
In the case of square lattice, the enlarged unit cell is the 
$\sqrt{2}\times\sqrt{2}$ one. The new reciprocal lattice vectors are 
$\vec{G}_1= 2\pi (1,-1)/a$,  
$\vec{G}_2= 2\pi (1,1)/a$, and the 
angle $\theta_k$ of Eq.(\ref{Mod:theta}) is 
\be
\theta_k = \frac{a}{2}(k_1+k_2) = k_xa.
\label{thetasquare}
\ee

In the case of the honeycomb lattice, the unit cell contains 
already two lattice sites. The energy is given by 
\ben
\epsilon_k^2 &=& \left[ 1 +  2\cos\left(\frac{3}{2}k_x a\right)
\cos\left(\frac{\sqrt{3}}{2}k_y a\right)\right]^2\\
&+& \left[2 \sin\left(\frac{3}{2}k_x a\right)
\cos\left(\frac{\sqrt{3}}{2}k_y a\right)\right]^2,
\een
and 
\[
\theta_k = \tan^{-1}\left(
\frac{
2 \sin\left(\frac{3}{2}k_x a\right)
\cos\left(\frac{\sqrt{3}}{2}k_y a\right) }
{
1+2 \cos\left(\frac{3}{2}k_x a\right)
\cos\left(\frac{\sqrt{3}}{2}k_y a\right) }
\right).
\]
The Brillouin zone is still honeycomb, with the $y$-axis one of its 
axes, and side equal to $4\pi/(3\sqrt{3}a)$.

\section{Ward Identity}
\label{App:Ward Identity}

Let us consider a generic Hamiltonian in the two sublattice 
representation 
\be
H = \sum_{R_1,R_2} \cc_{R_1} H_{R_1,R_2} \ca_{R_2},
\label{Ward:H}
\ee
where $H_{R_1,R_2}$ is a $2\times 2$ Hermitean matrix. The current 
operator 
\[
\vec{J}(R) = \sum_{R_1,R_2} \cc_{R_1} \vec{J}_{R_1,R_2}(R) \ca_{R_2},
\]
can be obtained by the continuity equation, leading to 
\be
\vec{\nabla}\cdot\vec{J}(R)  = 
i\sum_{R_1} \cc_R H_{R,R_1} \ca_{R_1}
-\cc_{R_1} H_{R_1,R} \ca_{R},
\label{B2}
\ee
being $\vec{\nabla}$ the discrete version of the differential operator. 
The long-wavelength expression for $\vec{J}(R)$ 
can be obtained by Fourier transformation, namely, through 
\[
i\sum_R \vec{q}\cdot\vec{J}(R) {\rm e}^{-i\vec{q}\cdot\vec{R}}
= 
i\sum_{R,R_1} \left(\cc_R H_{R,R_1} \ca_{R_1}
-\cc_{R_1} H_{R_1,R} \ca_{R}\right) {\rm e}^{-i\vec{q}\cdot\vec{R}},
\]
and expanding both sides in $q$, we get, for the linear term, 
\ben
&&i\sum_R \vec{q}\cdot\vec{J}(R)
= 
\sum_{R,R_1} \vec{q}\cdot\vec{R} \left(\cc_R H_{R,R_1} \ca_{R_1}
-\cc_{R_1} H_{R_1,R} \ca_{R}\right)\\
&& = 
\sum_{R,R_1} \vec{q}\cdot\left(\vec{R}_1-\vec{R}\right) 
\cc_{R_1} H_{R_1,R} \ca_{R},
\een
hence
\be
\vec{J}(R) = -i
\sum_{R_1} \left(\vec{R}_1-\vec{R}\right) 
\cc_{R_1} H_{R_1,R} \ca_{R}.
\label{Ward:current}
\ee
Let us define the correlation functions
\be
\chi_{\mu,i}(R,R';t,t_1,t_2)
= 
\langle T\left[ \cc_{R_1}(t) J^\mu_{R_1,R_2}(R)\ca_{R_2}(t)
\cc_{R_3}(t_1) J^i_{R_3,R_4}(R')\ca_{R_4}(t_2)\right]\rangle,
\label{B4}
\ee
where $\mu=0,1,2,3$, $J^0_{R_1,R_2}(R) = \delta_{RR_1}\delta_{RR_2}$ 
are the density matrix elements, and $J^i$, for $i=1,2,3$, are the 
matrix element components of the current. 
By the continuity equation, we find that 
\ben
\partial_t \chi_{0,i} + \partial_j \chi_{j,i} &=& i \sum_{R_1}
-\delta(t-t_1) Tr\left[ G(R_1,R;t_2-t)J^i_{R,R_1}(R')\right]\\
&& +\delta(t-t_2) Tr\left[ G(R,R_1;t-t_1)J^i_{R_1,R}(R')\right].
\een
If we integrate both sides by 
\[
\int dt\,dt_1\,dt_2, {\rm e}^{i(E+\omega)(t-t_1)}
{\rm e}^{iE(t_2-t)},
\]
at $\omega=0$ we find
\be
\partial_j \chi_{j,i}(R,R';E) = 
i\sum_{R_1}
Tr\left[ G(R,R_1;E)J^i_{R_1,R}(R')\right] -
Tr\left[ G(R_1,R;E)J^i_{R,R_1}(R')\right].
\ee
Using once more the continuity equation (\ref{B2}), we find 
\ben
&&\partial_j \partial'_i \chi_{j,i}(R,R';E) =
- Tr\left[ G(R,R';E)H_{R',R} + G(R',R;E)H_{R,R'}\right]\\
&& + \delta_{RR'}\sum_{R_1} 
Tr\left[ G(R,R_1;E)H_{R_1,R}\right] +
Tr\left[ G(R_1,R;E)H_{R,R_1}\right].
\een
By Fourier transform,
\ben
&&\sum_{RR'} \partial_j\partial'_i \chi_{j,i}(R,R';E)  
{\rm e}^{-i\vec{q}\cdot(\vec{R}-\vec{R'})} = 
\sum_{RR'} q_i q_j \chi_{j,i}(R,R';E) 
{\rm e}^{-i\vec{q}\cdot(\vec{R}-\vec{R'})}\\
&&\sum_{RR'}
\left(1-{\rm e}^{-i\vec{q}\cdot(\vec{R}-\vec{R'})}\right)
Tr\left[ G(R,R';E)H_{R',R} + G(R',R;E)H_{R,R'}\right].
\een
At small $q$, the above expression is 
\ben
&&
\sum_{RR'} q_i q_j \chi_{j,i}(R,R';E) \\
&& = \frac{1}{2} \sum_{RR'} 
q_iq_j \left(R_i-R'_i\right)\left(R_j-R'_j\right)
Tr\left[ G(R,R';E)H_{R',R} + G(R',R;E)H_{R,R'}\right],
\een
leading to 
\be
\sum_{RR'} \chi_{j,i}(R,R';E) 
= \sum_{RR'} \left(R_i-R'_i\right)\left(R_j-R'_j\right)
Tr\left[ G(R,R';E)H_{R',R} \right].
\label{Ward3}
\ee

\section{Longitudinal Modes}
\label{App:Longitudinal Modes}

As discussed in Section \ref{Longitudinal Fluctuations}, the expression 
of the $Q$-matrix which includes also the longitudinal modes is 
$Q_P(R)=Q(R) + S(R)$, where $Q(R)$ and $S(R)$ have been defined through 
(\ref{QPR}). The free action for these fields contains a local term, 
Eq.(\ref{SQ_Ploc}), and a non local one, Eq.(\ref{SQ_P}). The latter 
is 
\bea
&&-\frac{1}{4}\int dR\,dR'\, \Gamma(R-R') \,
Tr\left[ \Delta_R Q(R')\Delta_R Q(R')^\dagger \right.\nonumber\\
&&+\left. \Delta_R S(R')\Delta_R S(R')^\dagger
+2\Delta_R Q(R')\Delta_R S(R')^\dagger\right],
\label{delta:action}
\eea
where we have defined the operator 
\[
\Delta_R f(R')= f(R)-f(R') = \sum_{n=1}^\infty \frac{1}{n{\rm !}} 
\left(\vec{R}-\vec{R'}\right)^n\cdot \vec{\nabla}^n f(R').
\]
Let us apply this operator to $Q(R)$ and $S(R)$, keeping all terms which 
contains at most two derivatives which act to the transverse matrices $T$. 
We obtain
\bea
\Delta_R Q(R') &\simeq& 
\vec{\nabla}Q(R')\cdot \left(\vec{R}-\vec{R'}\right),
\label{DeltaQR}\\
\Delta_R S(R') &\simeq& \tilde{T}(R')^\dagger \Delta_R P(R') T(R')
\label{B.1}\\
&+&\left[ \left(\vec{\nabla}\tilde{T}(R')^\dagger\right) P(R) T(R')
+ \tilde{T}(R')^\dagger P(R) \left(\vec{\nabla}T(R')\right)\right]
\cdot \left(\vec{R}-\vec{R'}\right)
\label{B.2}\\
&+&\frac{1}{2}\left[
\left(\vec{\nabla}^2\tilde{T}(R')^\dagger\right) P(R) T(R')
+2\left(\vec{\nabla}\tilde{T}(R')^\dagger\right) P(R) 
\left(\vec{\nabla}T(R')\right)\right. \nonumber\\
&+& \left. \tilde{T}(R')^\dagger P(R) \left(\vec{\nabla}^2 T(R')\right)
\right]\cdot \left(\vec{R}-\vec{R'}\right)^2.
\label{B.3}
\eea
The term which is obtained by (\ref{B.1}) times its hermitean conjugate, 
together with the local piece (\ref{SQ_Ploc}) give the 
free action of the longitudinal modes
\be
S_0[P] = 
\frac{1}{V}\sum_q \frac{1}{2\omega_q} Tr\left[P_q P_q^\dagger\right].
\label{Appendix:SPfree}
\ee

The mixed terms, after defining $\vec{D}=T\vec{\nabla}T^\dagger$, 
give rise to the coupling between transverse 
and longitudinal modes
\bea
S[Q,P] &=&  
-\frac{1}{2}\int dR\, dR'\, \Gamma(R-R') 
\left(\vec{R}-\vec{R'}\right) 
Tr\left[\vec{D}(R')\left(P^\dagger(R')P(R) - 
P(R)^\dagger P(R')\right)
\right]  \label{Appendix:uno}\\
&-&\frac{1}{4}
\int dR\, dR'\, \Gamma(R-R') 
\left(\vec{R}-\vec{R'}\right)^2 \cdot\nonumber\\ 
&&Tr\left[\sigma_1 \vec{D}(R')\sigma_1 P(R) \vec{D}(R') P^\dagger(R') 
+ \sigma_1 \vec{D}(R')\sigma_1 P(R') \vec{D}(R') P^\dagger(R)
\right.\label{Appendix:tre}\\
&&\left. -\vec{D}(R')\vec{D}(R')\left(P^\dagger(R)P(R')
+ P^\dagger(R')P(R)\right) \right]\label{Appendix:due}\\
&-&\frac{1}{4}\int dR\, dR'\, \Gamma(R-R')
\left(\vec{R}-\vec{R'}\right) \cdot 
Tr\left[\vec{\nabla}Q(R')\Delta_R P(R')^\dagger + H.c.\right].
\label{Appendix:quattro}
\eea
The last term, (\ref{Appendix:quattro}), gives rise to higher gradient 
contributions, hence can be neglected.  

\subsection{Longitudinal propagators}
\label{Appendix:Longitudinal}
Before averaging over the longitudinal modes, we have to evaluate 
the longitudinal propagators. The matrix 
\[
P = \left(P_{0,0}s_0+P_{0,3}s_3\right)\sigma_0 
+ i \left(P_{3,1}s_1+P_{3,2}s_2\right)\sigma_3,
\]
has to satisfy $cP^t c^t = P$, and, in addition, all 
$P_\alpha=P_\alpha^\dagger$. For $\alpha=(0,0),(0,3),(3,1)$, by writing 
\[
P_\alpha = P^{(0)}_\alpha \tau_0 + i \vec{P}_\alpha\cdot\vec{\tau},
\]
we find that 
\[
P^{(0)}_\alpha,\vec{P}_\alpha\in {\cal R}e,\;\;
P^{(0)}_\alpha=\left( P^{(0)}_\alpha\right)^t,\;\;
\vec{P}_\alpha = -\left(\vec{P}_\alpha\right)^t.
\]
For $\alpha=(3,2)$, by writing 
\[
P_{3,2} = iP^{(0)}_{3,2}\tau_0 + \vec{P}_{3,2}\cdot\vec{\tau},
\]
we must impose 
\[
P^{(0)}_{3,2},\vec{P}_{3,2}\in {\cal R}e,\;\;
P^{(0)}_{3,2}=-\left( P^{(0)}_{3,2}\right)^t,\;\;
\vec{P}_{3,2} = \left(\vec{P}_{3,2}\right)^t.
\]
If $P^{(i)}_\alpha$ is a symmetric real matrix, its propagator is 
\be
\langle P^{(i)}_{\alpha,ab} P^{(i)}_{\alpha,cd} \rangle 
= \frac{G}{2} \left( \delta_{ad}\delta_{bc}+\delta_{ac}\delta_{bd}\right),
\label{Appendix:symm}
\ee
while if it is antisymmetric
\be
\langle P^{(i)}_{\alpha,ab} P^{(i)}_{\alpha,cd} \rangle 
= -\frac{G}{2} \left( \delta_{ad}\delta_{bc}-\delta_{ac}\delta_{bd}\right),
\label{Appendix:antisymm}
\ee
where $G_k=\omega_k/8V$, and $(a,b,c,d)$  are replica indices. 
By means of these propagators, we readily find that, if 
$M = M^{(0)}_i + i\vec{M}_i\cdot\vec{\tau}$, where $M^{(i)}$ are 
matrices in the replica space, then,
for $\alpha=(0,0),(0,3),(3,1)$, the following results hold 
\be
\langle P_\alpha M P_\alpha \rangle = -G c M^t c^t + 
2G Tr\left(M^{(0)}\right) = -G c M^t c^t + G Tr\left(M \right),
\ee
while, for $\alpha=(3,2)$, 
\be
\langle P_{3,2} M P_{3,2} \rangle = G c M^t c^t + 
2G Tr\left(M^{(0)}\right) = G c M^t c^t + G Tr\left(M \right).
\ee
More generally, 
\bea
\langle P Ms_i\sigma_j P^\dagger \rangle &=& 
-G c M^t c^t \left( s_i\sigma_j+ s_3s_i\sigma_js_3 + 
\sigma_3 s_1 s_i\sigma_js_1\sigma_3 
- \sigma_3 s_2 s_i\sigma_j s_2\sigma_3\right)\nonumber\\
&+& 
G Tr\left(M \right)\left( s_i\sigma_j + s_3s_i\sigma_js_3 + 
\sigma_3 s_1 s_i\sigma_js_1\sigma_3 
+ \sigma_3 s_2 s_i\sigma_j s_2\sigma_3\right),\\
\langle P Ms_i\sigma_j P \rangle &=& 
-G c M^t c^t \left( s_i\sigma_j+ s_3s_i\sigma_js_3 - 
\sigma_3 s_1 s_i\sigma_js_1\sigma_3 
+ \sigma_3 s_2 s_i\sigma_j s_2\sigma_3\right)\nonumber\\
&+& 
G Tr\left(M \right)\left( s_i\sigma_j + s_3s_i\sigma_js_3 - 
\sigma_3 s_1 s_i\sigma_js_1\sigma_3 
- \sigma_3 s_2 s_i\sigma_j s_2\sigma_3\right). 
\eea

For $j=0,3$ the above expression simplifies to 
\bea
\langle P Ms_i\sigma_j P^\dagger \rangle &=& 
-2G c \left(Ms_i\sigma_j \right)^t c^t 
+ 2G\sigma_j Tr\left(M s_i s_0 \right),\\
\langle P Ms_i\sigma_j P \rangle &=& 
-2G s_3 c \left(Ms_i\sigma_j \right)^t c^t s_3 
+ 2G\sigma_j Tr\left(M s_i s_3 \right),
\eea
while for $j=1,2$
\bea
\langle P Ms_i\sigma_j P^\dagger \rangle  
&=& -2G s_3 c \left(Ms_i\sigma_j \right)^t c^t s_3 
+ 2G\sigma_j Tr\left(M s_i s_3 \right),\\
\langle P Ms_i\sigma_j P \rangle &=& 
-2G c \left(Ms_i\sigma_j \right)^t c^t 
+ 2G\sigma_j Tr\left(M s_i s_0 \right).
\eea
 
\subsection{Averaging $S[Q,P]$}

We have now all what it is needed to proceed 
in the averaging over $P$. Here we just sketch the calculation, which is 
quite involved and requires the matrix properties of $\vec{D}$ which are 
determined in Appendix \ref{Appendix:RG}. We just remark that  
(\ref{Appendix:tre}) and (\ref{Appendix:due}) do not reproduce the 
correct stiffness term. Indeed, it is (\ref{Appendix:uno}), 
which contributes at second order, which cancels the additional terms 
and allows to express everything in terms of the matrix $Q$. 

By means of the previously calculated propagators of the longitudinal 
modes, we find that
\be 
\langle S[Q,P]\rangle_P = 
\frac{Y}{4d\Sigma^2} \int dR\, Tr\left[\vec{\nabla}Q(R)^\dagger
\vec{\nabla}Q(R)\right] + \frac{1}{8\Sigma^2} 
\left[Tr\left(Q(R)^\dagger\vec{\nabla}Q(R)\sigma_3\right) \right]^2,
\label{SQP_average}
\ee 
where 
\be
Y = \int dR\, \Gamma(R)\Gamma(R)^{-1} R^2.
\ee

We notice that the first term is a contribution to the Kubo 
conductivity and the second to the staggered density of states 
fluctuations of the diagrams where 
the current vertices are those proportional to the random hopping.

\subsection{Additional terms}
\label{App:Longitudinal Fluctuations}

The last class of corrections which generate new operators is 
obtained by expanding each Green's function in 
(\ref{Act:second}) at first order, $G_P = G_0 -iG_0PG_0$, leading to 
the term
\be
 \frac{1}{4}\langle Tr\left(GPGUGPGU\right)\rangle_P.
\label{P2}
\ee
For the sake of clarity, we will analyse this term only in the case of a 
cubic lattice, where the derivation is more straightforward. We will 
postpone a discussion about the general case at the end of the 
section. 

In the cubic lattice, according to Eq.(\ref{Simpsquare}), the 
electron-$Q$ coupling can be brought to a local one 
also in the diagonal basis. In this basis, $Q=Q_0 + iQ_1\sigma_1$, 
\be
P(R) = \left(P_{00}s_0 + P_{03}s_3\right)\sigma_0
+ i\left(P_{11}s_1 + P_{12}s_2\right)\sigma_1,
\label{PRsquare}
\ee
the unitary matrix 
\be
T(R) = \exp\left[ \frac{W_0(R)}{2}\sigma_0 + 
\frac{W_1(R)}{2}\sigma_1 \right],
\ee
where $W_1$ has the same form as $W_3$ defined by Eq.(\ref{Def:W}), 
and $\tilde{T}=\sigma_3 T \sigma_3$. Moreover, since $H_{RR'}=
\epsilon_{R-R'}\sigma_3$ in the diagonal basis, being 
$\epsilon_{R-R'}$ the Fourier transfom of $\epsilon_k$, 
the current operator 
which appears in the definition of $U_{RR'}$, Eq.(\ref{Act:U}), 
has only matrix elements $\vec{J}_k = \vec{\nabla}\epsilon_k\,\sigma_3$.

Once averaged over $P$, (\ref{P2}) gives, among other terms which 
correct the Kubo conductivity, a new term [see Eq.({Greenfunctiond})] 
\ben
&&\frac{1}{2}\sum_{a,b=1,2}\,\sum_{h=\pm}\, 
\sum_{pkq}\omega_{p-k}
Tr\left(U^{b,h;a,h}_q {\cal G}^{a,h}_{p+q} {\cal G}^{b,h}_p\right)
Tr\left(U^{a,h;b,h}_{-q}{\cal G}^{b,h}_k {\cal G}^{a,h}_{k+q}\right) \\
&&- Tr\left(U^{b,h;a,h}{\cal G}^{a,h}_{p+q}{\cal G}^{b,h}_p\right)
Tr\left(U^{-a,-h;-b,-h}{\cal G}^{-b,-h}_k {\cal G}^{-a,-h}_{k+q}\right),
\een
where the first piece derives from $P_0$ and the second from $P_1$.
The structure in the energy/sublattice indices can be 
shortly represented by
\[
\sum_{i=1,\dots,4} \Lambda_i \otimes \Lambda_i =
\frac{1}{4} \left( \sigma_2\otimes\sigma_2 + s_3\sigma_1\otimes s_3\sigma_1
+ \sigma_3\otimes\sigma_3 + s_3\otimes s_3 \right), 
\]
so that the above term can be written, at small $q$, as 
\be
\frac{1}{2}\sum_{i=1,\dots,4} 
\sum_{pkq}\omega_{p-k}
Tr\left(U_q {\cal G}_{p}\Lambda_i {\cal G}_p\right)
Tr\left(U_{-q}{\cal G}_k\Lambda_i {\cal G}_{k}\right) 
\ee
We remind that 
\[
{\cal G}_p U_q {\cal G}_{p} = -i
{\cal G}_p \vec{D}_q \cdot \vec{J}_{p} {\cal G}_{p},
\]
where $\vec{D}_q$ is the Fourier transform of $\tilde{T}(R)
\vec{\nabla}\tilde{T}(R)^{-1}$. We notice that only the diagonal 
component in energy of $\vec{D}$ enters. Moreover, for the diagonal matrices  
$s_j=s_0,s_3$, 
since $\vec{D}=\vec{D}_0\sigma_0 + \vec{D}_1\sigma_1$, it derives that, 
for $i=0,1$, the following equality holds   
$Tr \vec{D}s_j\sigma_i = -
Tr \left(\vec{\nabla}\tilde{W} s_j\sigma_i/2\right)$. Therefore,  
just $W_1\sigma_1$ contributes. 
By means of (\ref{Greenfunctiond}), for any $\Lambda$'s we have
\bea
&&-iTr\left(\vec{D}_{1,q}\sigma_1\cdot \vec{J}_p {\cal G}(p) 
\Lambda_i {\cal G}(p)\right)
= + \vec{\nabla}\epsilon_p \cdot Tr\left(\vec{D}_{1,q} s_1{\cal G}(p)s_1 
\sigma_2 \Lambda_i {\cal G}(p)\right)\nonumber\\
&&= \vec{\nabla}\epsilon_p \cdot Tr\left[\vec{D}_{1,q}
\left(G_3^2 \sigma_3\sigma_2\Lambda_i\sigma_3
-G_0^2 s_3\sigma_2\Lambda_i s_3 
+ G_3 G_0 s_3 [\sigma_3,\sigma_2\Lambda_i]\right)\right].
\eea
Only for $\Lambda_1=\sigma_2/4$ the trace over the $\sigma$'s 
is finite, leading to 
\be
2\left(G_3^2 - G_0^2\right)\vec{\nabla}\epsilon_p 
\cdot Tr\left(\vec{D}_{1,q}\right).
\ee

In conclusion, going back to the original sublattice representation, 
and defining  
\be
\Pi^{(0)} = \frac{1}{4\pi V^2 d}\sum_{pk}  
\vec{\nabla}\epsilon_k \cdot \vec{\nabla}\epsilon_p \,
\omega_{p-k} Tr\left(\sigma_3G^+_{p}\sigma_3G^+_{p}\right)
Tr\left(\sigma_3 G^+_{k}\sigma_3 G^+_{k}\right),
\label{Pi:definition}
\ee
we obtain the following explicit expression of (\ref{P2}) 
[notice that $Q= Q_0\sigma_0 + iQ_1\sigma_1 \to Q_0\sigma_0 + i Q_3\sigma_3$ 
in the sublattice basis]
\be
\frac{2\pi}{8\cdot 32  \Sigma^4} \Pi^{(0)}
\int dR\,
Tr\left[Q^\dagger(R)\vec{\nabla}Q(R)\sigma_3\right]
\cdot Tr\left[Q^\dagger(R)\vec{\nabla}Q(R)\sigma_3\right].
\label{App:Anomalo}
\ee
This term, which derives from the regular current vertices, 
has the same form as the second term in
(\ref{SQP_average}), which, on the contrary, is due to the random 
current vertices. 
Therefore, the coupling constant $\Pi$ which appears in the 
final expression (\ref{Anomalo}) is the sum of both  
terms.

The symmetry of the operator (\ref{App:Anomalo}), which, due to the trace, 
involves the Nambu, energy and sublattice components $\tau_0$, $s_0$ and 
$\sigma_3$, respectively, suggests that the 
prefactor $\Pi$ represents the fluctuations of the  
staggered density of states, as discussed in section 
\ref{Physical Meaning of Pi}. 
In the case of a generic bipartite lattice, 
we do expect an analogous term to appear, 
because the staggered-density of states fluctuations still acquire singular 
contributions, and the 
operator is not forbidden by the symmetry properties of the 
$Q$-matrix. 
The reason why we decided to show only the case of cubic lattices is that 
the distinction  
between the longitudinal from the transverse modes is a bit ambiguous 
at large momenta, where the both are in a sense massive. 
This is not a problem for cubic lattices, where one can show that 
the large momentum components of the transverse modes do not 
contribute, hence (\ref{App:Anomalo}) exhausts the whole contribution. 
On the contrary, in other cases, we do have to keep into account 
the contribution of the small-wavelength transverse modes to recover 
the full expression. 
This makes the calculations more involved 
than in the case of cubic lattices.

\section{Explicit derivation of the RG equations}
\label{Appendix:RG}

In this Appendix we outline the derivation of the RG equations.  
Before that, it is convenient to list some useful results.

\subsection{Averages}

Besides the propagators, we will also need the explicit expression of 
particular averages which enter in the derivation of the RG equations. 
The following results hold:
\bea
\langle B_{ab}P_{bc}B_{cd}^\dagger\rangle &=& 
 4D(k)\delta_{ad} Tr\left(P_0\right),\label{BBdag}\\
\langle B_{ab}P_{bc}B_{cd}\rangle &=& 
 -2D(k)P^\dagger_{ad},\label{BB}
\eea
where $P=P_0+i\vec{P}\cdot\vec{\tau}$ is a quaternion real matrix. 

In addition,
\bea
\langle A_{ab}P_{bc}A_{cd}\rangle &=& 
-2D(k)P^\dagger_{ad} + 4D(K)\delta_{ad}Tr\left(P_0\right)\nonumber\\
&-& D(k)\frac{\Pi}{\sigma_{xx}+\Pi m} P_{ad},\label{AA}\\
\langle C_{ab}P_{bc}C_{cd}\rangle &=& 
-2D(k)P^\dagger_{ad} + 4D(K)\delta_{ad}Tr\left(P_0\right)\nonumber\\
&-& D(k)\frac{\Pi}{\sigma_{xx}+\Pi m} P_{ad},\label{CC}\\
\langle A_{ab}P_{bc}C_{cd}\rangle &=& 
- D(k)\frac{\Pi}{\sigma_{xx}+\Pi m} P_{ad}.
\label{AC}
\eea

For instance, if $P=\hat{I}$, then 
\bea
\langle BB^\dagger \rangle 
&=& 4D(k)m \hat{I}\label{BBdagI}\\
\langle BB \rangle 
&=& -2D(k)\hat{I}\label{BBI}\\
\langle AA \rangle 
&=& \left[ -2D(k) + 4D(k)m - D(k)\frac{\Pi}{\sigma_{xx}+\Pi m}
\right]\hat{I}\label{AAI}\\
\langle CC \rangle 
&=& \left[ -2D(k) + 4D(k)m - D(k)\frac{\Pi}{\sigma_{xx}+\Pi m}
\right]\hat{I}\label{CCI}\\
\langle AC \rangle 
&=& - D(k)\frac{\Pi}{\sigma_{xx}+\Pi m} \hat{I}.\label{ACI}
\eea

\subsection{RG equations}
First of all, we need to know the quaternion structure of the matrix 
$D = T \nabla T^\dagger$. Since 
\[
D = - D^\dagger,\;\;\; c D^t c^t = -\sigma_1 D \sigma_1,
\]
we can write $D=D_0\sigma_0 + D_3\sigma_3$, where in the 
$\pm$-space
\be
D_0 = 
\left(
\begin{array}{cc}
A_0 & B_0\\
-B_0^\dagger & C_0\\
\end{array}
\right),
\label{App:D0}
\ee
and 
\be
D_3 = 
i\left(
\begin{array}{cc}
A_3 & B_3\\
B_3^\dagger & C_3\\
\end{array}
\right).
\label{App:D3}
\ee
We can write each of the above matrices in quaternion form, 
$P = P^{(0)}\tau_0 + i\vec{P}\cdot\vec{\tau}$, where 
$P^{(0)},\vec{P}\in{\cal R}e$, but, in addition, we must impose that 
\[
\begin{array}{cccc}
\left(A_0^{(0)}\right)^t = - A_0^{(0)} &
\vec{A}_0^t = \vec{A}_0 &
\left(C_0^{(0)}\right)^t = - C_0^{(0)} &
\vec{C}_0^t = \vec{C}_0 \\
\left(A_3^{(0)}\right)^t =  A_3^{(0)} &
\vec{A}_3^t = -\vec{A}_3 &
\left(C_3^{(0)}\right)^t =  C_3^{(0)} &
\vec{C}_3^t = -\vec{C}_3 \\
\end{array}
\]
By making use of the above properties, and by means of the Eqs.(\ref{BB}), 
(\ref{BBdag}), (\ref{AA}), (\ref{CC}) and (\ref{AC}), after defining 
\[
L = \int_{\Lambda/s}^\Lambda \frac{d^2 k}{(2\pi)^2} D(k),
\]
we get the following results
\bea
&&\langle W_f D_0 W_f \rangle_f = L_A \Gamma D_0 \nonumber \\
&& +  
2L_B \left(
\begin{array}{cc}
0 & B_0\\
-B_0^\dagger & 0\\
\end{array}
\right) 
-2L_A \left(
\begin{array}{cc}
A_0 & 0\\
0 & C_0\\
\end{array}
\right),
\label{App:WD0W}
\eea
where $\Gamma = \Pi/(\sigma_{xx}+m\Pi)$, and 
\bea
&&\langle W_f D_3 W_f \rangle_f = L_A \Gamma D_3 \nonumber \\ 
&& + 2L_A i\left(
\begin{array}{cc}
A_3 & 0\\
0 & C_3\\
\end{array}
\right) 
-2L_B i\left(
\begin{array}{cc}
0 & B_3\\
B_3^\dagger & 0\\
\end{array}
\right) \nonumber\\
&&-4L \hat{I} Tr\left(iA_3^{(0)}+iC_3^{(0)}\right).
\label{App:WD3W}
\eea
For further convenience, when useful, we have labelled $L_B$ 
the propagator of $B_f$, and $L_A$ the ones of 
$A_f$ and $C_f$.  The next useful result is 
\be
\langle W_f W_f \rangle_f = 
\left(-4L_B m -4L_A m + 2L_A + L_A\Gamma\right)\hat{I}.
\label{App:WW}
\ee
Through (\ref{App:WD0W}) and (\ref{App:WD3W}) we therefore get 
\bea 
\langle F_1 \rangle_f = 
&=& \frac{2\pi\sigma_{xx}}{32\Sigma^2} \int dR\, 
-2 \langle Tr\left[ \vec{D}\sigma_1 Q_{sp} W_f \vec{D} W_f Q_{sp}
\sigma_1\right]\rangle_f
\nonumber\\ 
&&+2 \langle Tr\left[ \vec{D}\sigma_1 Q_{sp} \vec{D} Q_{sp} 
\tilde{W}_f^2
\sigma_1\right]\rangle_f \nonumber\\
&=&
\frac{2\pi\sigma_{xx}}{32} \int dR\,
4\left(L_B-L_A+2L_Am+2L_Bm\right) 
Tr\left(
\begin{array}{cc}
0 & B_0\\
-B_0^\dagger & 0\\
\end{array}
\right)^2 
\label{App:RGuno}\\
&+& 4\left(2L_Am+2L_Bm\right)
\left(
\begin{array}{cc}
iA_3 & 0\\
0 & iC_3\\
\end{array}
\right)^2 
\label{App:RGdue}\\
&-& 2 L \left[Tr\left(D\sigma_3\right)\right]^2
\label{App:RGtre} \\
&-&4\left(2L_Am+2L_Bm+2L_A\right)
\left(
\begin{array}{cc}
A_0 & 0\\
0 & C_0\\
\end{array}
\right)^2
\label{App:RGquattro}\\
&-& 4\left(2L_Am+2L_Bm+L_A+L_B\right)
\left(
\begin{array}{cc}
0 & iB_3\\
iB_3^\dagger & 0\\
\end{array}
\right)^2.
\label{RGcinque}
\eea

The calculation of $\langle F_2^2 \rangle_f$ is more involved, since 
one needs the average of four $W_f$'s. For sake of lengthy, we just 
quote the final result that such a term cancels  
(\ref{App:RGquattro}) and (\ref{RGcinque}). 
We next notice that 
\ben
&&\frac{1}{\Sigma^2}Tr\left[\nabla Q^\dagger \nabla Q\right]
=
2Tr\left[Ds_3\sigma_1 D s_3 \sigma_1 - DD\right]\\
&&=
-4 Tr\left(
\begin{array}{cc}
0 & B_0\\
-B_0^\dagger & 0\\
\end{array}
\right)^2 
-4 Tr\left(
\begin{array}{cc}
iA_3 & 0\\
0 & iC_3\\
\end{array}
\right)^2 ,
\een
and that 
\[
Tr\left[Q^\dagger \nabla Q \sigma_3\right]
= 2 \Sigma^2 Tr\left(D\sigma_3\right).
\]
Therefore, for $L_A=L_B$,  
\bea
&&\langle F_1 -\frac{1}{2}F_2^2\rangle_f \nonumber\\
&& =
-4Lm \frac{2\pi\sigma_{xx}}{32\Sigma^2} \int dR\,
Tr\left[\nabla Q^\dagger \nabla Q\right]
\nonumber\\
&& -\frac{1}{2} L \frac{2\pi\sigma_{xx}}{32\Sigma^4}
\left[Tr\left(Q^\dagger \nabla Q \sigma_3\right)\right]^2.
\label{App:RG}
\eea
In the cases in which the $A$ and $C$ modes are gaped ($L_A=0$, 
and no $D_3$), we obtain the 
standard result 
\be
\langle F_1 -\frac{1}{2}F_2^2\rangle_f =
-\left(2Lm+L\right) \frac{2\pi\sigma_{xx}}{32\Sigma^2} \int dR\,
Tr\left[\nabla Q^\dagger \nabla Q\right].
\label{App:RGnoA}
\ee

\section{Gade and Wegner's proof of the vanishing $\beta$-function}
\label{App:Symmetries}

The equations from (\ref{Def:W}) to 
(\ref{AiCi:condition}) imply that $W_3$ is not a traceless 
matrix. Indeed, we can alternatively write $W$ as
\be
W = W' + \frac{1}{4m}Tr\left(W_3\right)\sigma_3
\equiv W' + i\phi\sigma_3,   
\label{App:W}
\ee
with $W'=W_0\sigma_0 + W'_3\sigma_3$, now being $W'_3$ a traceless matrix. 
Since $\sigma_3$ commutes with $W'$, this means that 
\be
T(R)^2 = {\rm e}^{W(R)} = {\rm e}^{i\phi(R)\sigma_3} {\rm e}^{W'(R)} 
\equiv {\rm e}^{i\phi(R)\sigma_3} V(R),
\label{App:U}
\ee
which also defines the matrix field $V(R)$. 
By means of this parametrization, the non-linear 
$\sigma$-model (\ref{Summary:NLSM}) can also be written as 
\bea
S[T] = S[V,\phi] &=& \frac{2\pi\sigma_{xx}}{32} \int dR\; 
Tr\left[ \vec{\nabla}V(R)^{-1} \cdot \vec{\nabla}V(R)\right]
\nonumber\\
&+& \frac{m\pi}{2}\left(\sigma_{xx}+m\Pi\right)
\int dR\; \vec{\nabla}\phi(R)\cdot\vec{\nabla}\phi(R).
\label{App:NLSM}
\eea
Therefore the action of $V$ is distinct from that of $\phi$, and the 
latter, being a phase, is gaussian. This implies that the 
combination $\sigma_{xx}+m\Pi$ is not renormalized and scales with its 
bare dimension $\epsilon$, for any number of replicas. In turns, 
it means that, in the zero replica limit, it is $\sigma_{xx}$ which 
is not renormalized! This is completely equivalent to the nice proof 
given by Gade and Wegner\cite{W&G} that the quantum corrections to the 
$\beta$-function of the conductance of a ${\rm U}(N)/{\rm SO}(N)$ model 
vanish at all orders in the $N\to 0$ limit. 

The other important result 
concerns the renormalization of an operator 
\[
T^{2q} = {\rm e}^{iq\phi\sigma_3} V^q.
\]
Within RG, 
\be
{\rm e}^{iq\phi\sigma_3} \to \frac{t}{2} q^2 \ln s 
\left(\frac{t}{c+mt} - \frac{1}{m}\right) {\rm e}^{iq\phi\sigma_3}.
\label{App:RGphi}
\ee
The second term, which is singular in the $m\to 0$ limit, 
has to be canceled  
by the one-loop renormalization of $V^q$. 
Gade and Wegner showed that this cancellation holds for any $q$. 
Furthermore, they argued that, apart from the one-loop correction, 
the renormalization of $V^q$ does not contain any other singular term 
in the $m\to 0$ limit.  This, as they pointed out, has very important 
consequences. In $2d$, $t$ does not scale, while $c$ goes to zero. 
Therefore the term which dominates the renormalization of 
$T^{2q}$ for $m=0$ is just the first term in the right hand side 
of (\ref{App:RGphi}). This argument implies that the one-loop correction, 
which we have derived for the density of states ($q=1$ case), 
is sufficient to identify the correct asymptotic behavior.  

To conclude, let us discuss 
more in detail the origin of this gaussian field $\phi$. 
In the Grassmann variable path-integral representation, the action for the 
particle-hole symmetric model at $E=\omega=0$ 
\[
S = \sum_{RR'} \overline{\Psi}_R H_{RR'}\Psi_{R'},
\]
posseses a simple abelian gauge-like symmetry 
\be
\Psi \to {\rm e}^{i\phi\sigma_3}\Psi,
\label{Abelian}
\ee
because $\{ \sigma_3, H_{RR'} \}=0$. It is just this symmetry which causes the 
appearance of the gaussian part of the non-linear $\sigma$-model. 
Notice that this symmetry implies a particle-hole symmetric 
Hamiltonian, which is invariant under the transformation 
$\ca_{1,R}\to \cc_{1,R}$ but $\ca_{2,R}\to -\cc_{2,R}$. In fact,  
$\{ \sigma_3, H_{RR'} \}=0$ also means that 
$\{ \tau_1\sigma_3,H_{RR'}\}=0$, being $H_{RR'}\propto\tau_0$. 
The interesting fact is that (\ref{Abelian}) is not a 
symmetry of the fermion operators. 
Indeed,  under this transformation, 
$c_R \to {\rm e}^{i\phi\sigma_3} c_R$, but 
$\overline{c}_R \to {\rm e}^{i\phi\sigma_3} \overline{c}_R$, and not  
$\overline{c}_R \to {\rm e}^{-i\phi\sigma_3} \overline{c}_R$ as we would 
expect if $c_R$ and $\overline{c}_R$ had to be identified with the 
operators $\ca_R$ and $\cc_R$. Finally, we notice that if, besides 
$\sigma_3$, the Hamiltonian commutes with another Pauli matrix 
(as it can be the case for specifically built particle-hole 
symmetric models), 
the above gauge symmetry would be non abelian, hence spoiling all 
peculiar properties which we have shown to occur. 
Indeed, in the last case, the system can be mapped into a 
standard localization model with an additional sublattice index.

\end{document}